\documentclass[10pt,journal,compsoc]{IEEEtran}
\usepackage{amsmath,amsfonts}
\usepackage{algorithmic}
\usepackage{algorithm}
\usepackage{array}
\usepackage[caption=false,font=normalsize,labelfont=sf,textfont=sf]{subfig}
\usepackage{textcomp}
\usepackage{stfloats}
\usepackage{url}
\usepackage{verbatim}
\usepackage{graphicx}
\usepackage{cite}

\usepackage{algorithm}
\usepackage{algorithmic}
\usepackage{amsmath}
\usepackage{booktabs}
\usepackage{array,caption}
\usepackage{tabularx}
\usepackage{lipsum} % 用于生成示例文本
\usepackage{multirow}
\usepackage{graphicx}
\usepackage{color}
\usepackage{amsfonts}
\usepackage{relsize}
\usepackage{caption}
\begin{document}
\title{Explainable Session-based Recommendation via Path Reasoning}

\author{Yang Cao, Shuo Shang, Jun Wang, and Wei Zhang
\IEEEcompsocitemizethanks{
\IEEEcompsocthanksitem Yang Cao, Jun Wang and Wei Zhang are with the School of Computer Science and Technology, East China Normal University, Shanghai, China. E-mail: \{caoyang99775, wongjun, zhangwei.thu2011\}@gmail.com.
\IEEEcompsocthanksitem Shuo Shang  is with School of Computer Science and Engineering and Shenzhen Institute for Advanced Study University of Electronic Science and Technology of China, Chengdu, China. E-mail: {jedi.shang}@gmail.com.
}
}

        % <-this % stops a space


\IEEEtitleabstractindextext{%
\begin{abstract}
This paper explores providing explainability for session-based recommendation (SR) by path reasoning. Current SR models emphasize accuracy but lack explainability, while traditional path reasoning prioritizes knowledge graph exploration, ignoring sequential patterns present in the session history. Therefore, we propose a generalized hierarchical reinforcement learning framework for SR, which improves the explainability of existing SR models via Path Reasoning, namely PR4SR. Considering the different importance of items to the session, we design the session-level agent to select the items in the session as the starting point for path reasoning and the path-level agent to perform path reasoning. In particular, we design a multi-target reward mechanism to adapt to the skip behaviors of sequential patterns in SR, and introduce path midpoint reward to enhance the exploration efficiency in knowledge graphs. To improve the completeness of the knowledge graph and to diversify the paths of explanation, we incorporate extracted feature information from images into the knowledge graph. We instantiate PR4SR in five state-of-the-art SR models (i.e., GRU4REC, NARM, GCSAN, SR-GNN, SASRec) and compare it with other explainable SR frameworks, to demonstrate the effectiveness of PR4SR for recommendation and explanation tasks through extensive experiments with these approaches on four datasets.
\end{abstract}

}

\maketitle

\section{Introduction}
\IEEEPARstart{B}{oth} session-based recommendation (SR) and explainable recommendation have gained great attention in recent years. However, most of the current SR models focus on the problem of how to improve the accuracy of recommendations, while neglecting the process of providing explainability. A large number of research works design different model structures to capture user preference information and model sequential patterns, such as recurrent neural network~\cite{rs_gru4rec}, attention mechanism~\cite{rs_narm,rs_sasrec,rs_stamp} and graph neural network~\cite{rs_gcsan,rs_srgnn}. However, none of these structures are explainable. And a series of studies ~\cite{rl_adac,rl_sentiment,fu2020fairness,xian2020cafe,lu2023user,liu2023social,lyu2022knowledge,wei2023rule} have shown that an explainable recommendation process can improve the persuasiveness and trustworthiness of recommendation systems. Therefore, some generalizable explainability frameworks are needed to improve the explainability of existing SR models.

\graphicspath{{figs/}}
\begin{figure}[t!]
    \centering
    \includegraphics[width=8cm]{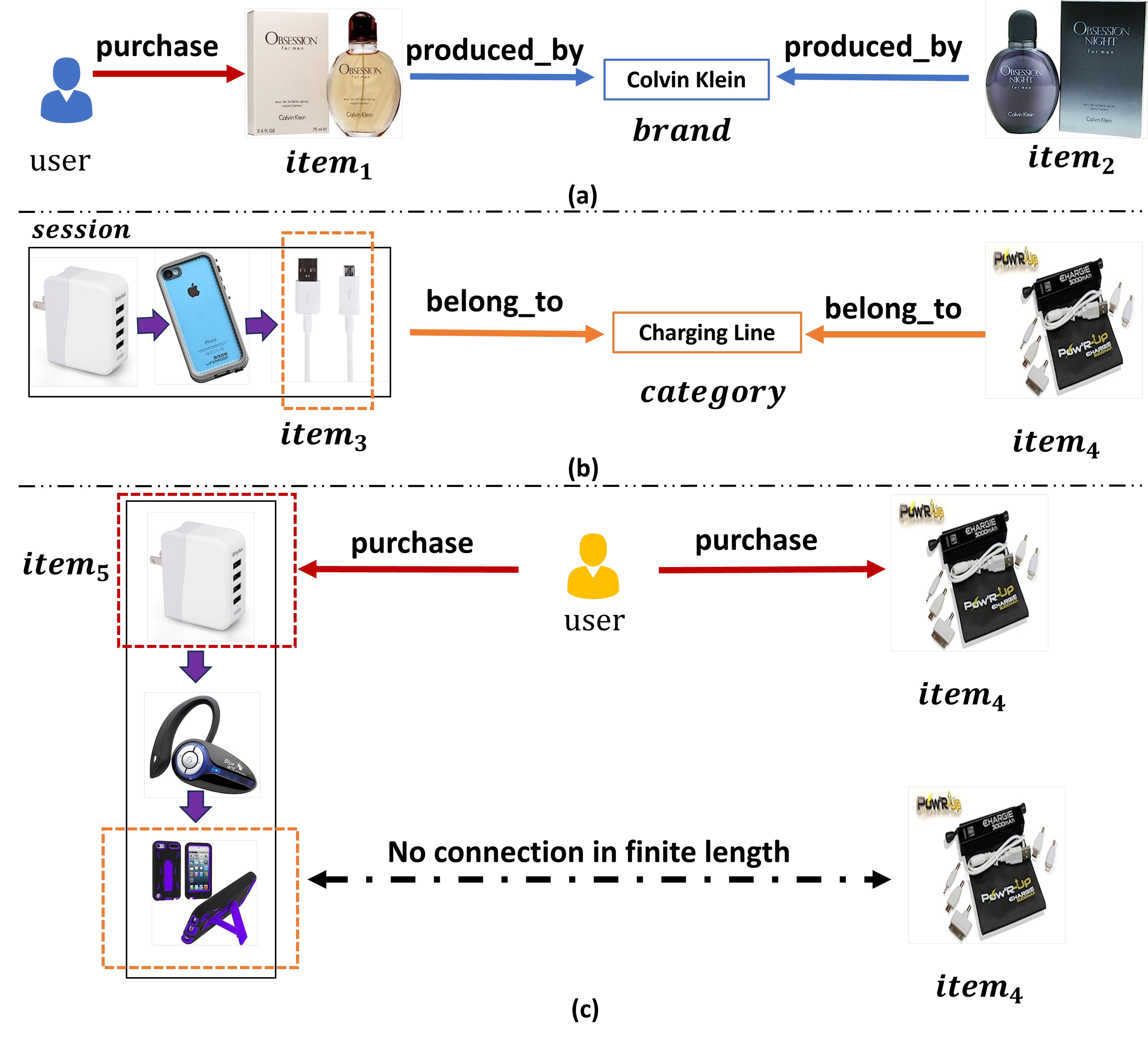}
    \caption{Three examples of path reasoning: (a) starting with a user, (b) starting with the last item, (c) starting with the most relevant item.}
    \label{fig1}
\end{figure}

Path reasoning~\cite{rl_multi,lin2018multi,sun2018recurrent,geng2022path} methods provide explainable information for recommendations by exploring paths between entities in the knowledge graph. In order to avoid enumerating possible reasoning paths in large-scale knowledge graphs~\cite{fu2020fairness,ma2019jointly,wang2019explainable,zhu2020knowledge}, recent research work combines reinforcement learning methods with path reasoning to improve the efficiency of exploring paths in knowledge graphs by designing suitable reward functions~\cite{icarte2022reward}. However, the current reward function only focuses on finding valid target points in the knowledge graph and exploring the transfer process between neighboring items in a session, but fails to take into account the skip behaviors of sequential patterns~\cite{rs_caser} by setting corresponding reward values for them.

Currently path reasoning methods can be divided into two categories. 
The first part~\cite{rl_pgpr,rl_multi} integrates rich heterogeneous information from the knowledge graph into the recommendation process, utilizing a path-level agent learning to navigate from a given user entity to potential items of interest to the user. Fig.~\ref{fig1} (a) illustrates a path of inference, $ user\xrightarrow{\text{purchase}} item_1 \xrightarrow{\text{produced\_by}} Colvin \enspace Klein \xleftarrow{\text{produced\_by}} item_2 $, which indicates that the reason for recommending $item_2$ to the user is that the user has purchased $item_1$ before, and the recommended item $item_2$ belongs to the same brand as $item_1$. However, this part of the work does not integrate session information as well as the design of the reward function only considers how to explore efficiently in the knowledge graph. 
The other part~\cite{rs_reks} considers combining session information and heterogeneous information in the knowledge graph to provide an explainable framework for SR, which fuses session information and heterogeneous information into the state representation for reinforcement learning and selects the last item in the session as the starting point for the path reasoning, and then plans an explainable path from the interacted item entity to the recommended item entity. For example in Fig.~\ref{fig1} (b), the last item in the session is chosen as the starting point for path reasoning, and a path, $ item_3 \xrightarrow{\text{belong\_to}} Charging \enspace Line \xleftarrow{\text{belong\_to}} item_4 $, is inferred from the knowledge graph, which indicates that the reason for predicting $item_4$ as the next is that both $item_4$ and the last $item_3$ in the session belong to the charging line category. Although this type of approach has taken session information into account, the design of the reward function still does not cover the skip behaviors of sequential patterns.

Moreover, some studies of SR have shown that the last item in a session does not represent the entire session sequence of interest. Caser~\cite{rs_caser} captures two sequence patterns in a sequence with horizontal filters, e.g., for a session [Airport, Hotel, Fast Food, Restaurant], the next behavior may be Greta Wall due to (Airport, Hotel) or Bar due to (Fast Food, Restaurant). NARM~\cite{rs_narm} finds that the most important items to the session appear at the beginning or in the middle of the session in some cases. These studies also suggest that simply choosing the last item in a session as a starting point for path reasoning is not sufficient to provide an explainable framework for SR. But existing explainable SR frameworks~\cite{rs_reks} only consider using the last item in the session as the starting point for path reasoning, which may lead to a failure to provide explainable paths for SR. For example in Fig.~\ref{fig1} (c),  if the last item is chosen as the starting point of path reasoning, because the last item and $item_4$ are the cell phone holder and cell phone charging cable respectively and the connection between these two items is not strong in the knowledge graph, there does not exist a finite-length path connecting these two items in the knowledge graph; if the first item is chosen, it can be extrapolated in finite-length path extrapolation because the first item $item_5$ is the charging plug and $item\_4$ is the charging cable, which are more closely related to each other. There exists such a path in the knowledge graph, $ item_5 \xleftarrow{\text{purchase}} user \xrightarrow{\text{purchase}} item_4 $, indicating that the exists user purchased both items.

Therefore, although the current research work demonstrates that it is possible to provide an explainable process for recommendation through path reasoning, the current path reasoning approach is not suitable for generalization in the scenario of SR. 
The main problem lies in the fact that previous research on path reasoning has only considered how to perform path reasoning in knowledge graphs, whereas for session recommendation scenarios, the impact of different items in a session on path reasoning should also be explored, as well as the consideration of designing appropriate reward functions for skip behaviors of sequential patterns in session scenarios.

In addition, a series of related study~\cite{rl_pgpr,rl_adac} demonstrate that the shorter the path length produced by path reasoning, the less time the reasoning takes as well as the more acceptable the explainable paths are. So that under the constraint of finite length, the closer the associations between the entities in the knowledge graph are, the higher the likelihood of exploring reasonable paths through path reasoning. In e-commerce scenarios, the images of products contain some feature information of the products, which can be used as product attributes to supplement the knowledge graph. However, in the current methods of path reasoning to construct knowledge graphs, the image information of products is not taken into account, which leads to less attribute information of products and fewer interconnections in the constructed knowledge graphs.

To address these challenges, we propose a generalized hierarchical reinforcement learning framework for SR, which improves the explainability of existing SR models through Path Reasoning (denoted as PR4SR). 
We design the session-level agent to select important items from the session as the starting point for path reasoning. We then make path reasoning to provide explainable paths by the proposed path-level agent.
In particular, we design a multi-target reward mechanism to adapt to the skip behaviors of sequential patterns in SR, and introduce path midpoint reward to enhance the exploration efficiency in knowledge graphs.
In order to make the interconnections of entities in the knowledge graph closer and increases the diversity of explainable paths, we have extracted the product feature information in the image by using image recognition method, 
and add them to the knowledge graph in the form of entities.

In summary, our contributions lie in three aspects:
\begin{itemize}
\item PR4SR is the first path reasoning approach that uses hierarchical reinforcement learning to provide a generalized and explainable framework for SR, where the session-level agent selects the important items in the session as the starting point for path reasoning and the path-level agent performs path reasoning in the knowledge graph. 

\item PR4SR is the first path reasoning approach that combines the skip behaviors of sequential patterns in SR into the design of the reward mechanism. The reward function is also designed to improve the exploration efficiency by considering the distance from the midpoint of the path to the goal points.

\item Comparing with the method of constructing product knowledge graphs in traditional path reasoning, a new method of constructing product knowledge graphs incorporating product picture features is designed, which improves the correlation between entities and the diversity of explainable path forms.

\item PR4SR generalizes well and can be combined with existing unexplainable SR models to accomplish both recommendation and explainability tasks at the same time. We compare PR4SR with state-of-the-art methods on four public datasets. The results show that PR4SR improves recommendation accuracy and model explainability for unexplainable SR models. We also conduct a comprehensive ablation study to analyze the contributions of key components. We further illustrate that the explainability of PR4SR outperforms other explainable SR frameworks through a user survey.
\end{itemize}

\section{Related Work}
\subsection{Session-based recommendation}
Early work has used MF~\cite{rs_mf} to learn users' overall interest preferences and MC~\cite{rs_mc} to learn the dependencies between items. FPMC~\cite{rs_fpmc} introduces a personalized transfer matrix based on Markov chains, which can capture both temporal information and long-term user preference information, and a Matrix Decomposition model is introduced to solve the sparsity problem of the transfer matrix.

The core idea of this type of models is to mine the potential relationship between users and items by decomposing or completing the user-item scoring matrix for personalized recommendation. Their workings are not easily understood by the general user and lack intuitiveness and explainability.

Deep neural network techniques are now being applied to SR models. In order to capture the user's interest at different points in time, GRU4Rec~\cite{rs_gru4rec} captures the interest information through the GRU~\cite{GRU} layer to improve the accuracy of recommendations.
To combine the user's interest in the current session and the sequence information of the session, NARM~\cite{rs_narm} proposes a hybrid state encoder with an attention mechanism.
Focusing on the fact that previous models do not consider the impact of the user's current action on the next action, STAMP~\cite{rs_stamp} proposes a short-term attention/memory priority model.
SASRec~\cite{rs_sasrec} finds that MC performs better in sparse datasets while RNN~\cite{RNN} performs better in dense datasets, thus proposing a self-attention based sequential model to balance MC and RNN.
SR-GNN~\cite{rs_srgnn} argues that previous research approaches ignore complex transformations between items, thus uses GNN~\cite{GNN} to capture complex transformations of items in graph-structured data. 
GCSAN~\cite{rs_gcsan} also focuses on the importance of dependencies between items and proposes a graph contextualized self-attention model.

Although these models improve the accuracy of recommendations by designing different network structures, the study of model explainability has been neglected, which may reduce the user's satisfaction with the whole recommender system.

\subsection{Path reasoning approaches}
Path reasoning approaches explicitly model the paths between entities over a knowledge graph for recommendation, and improve the efficiency of exploration in the knowledge graph through reinforcement learning methods. Such current work on path reasoning combining reinforcement learning and knowledge graphs can be categorized into two groups.

One of them uses the representation of the knowledge graph as the state representation for reinforcement learning and the user entity as a starting point for path reasoning. 
PGPR~\cite{rl_pgpr} first emphasizes the importance of incorporating knowledge graphs into recommendations to provide reasoning and explainable paths for recommendations, and proposes for the first time reasoning of explainable paths in the knowledge graph by means of reinforcement learning. 
ADAC~\cite{rl_adac} extracts path demonstrations from the knowledge graph as guided paths for reinforcement learning exploration, which improves the speed of model fitting.
TPRec~\cite{rl_time} uses GMM~\cite{ml_gmm} to cluster the purchase relation and incorporates the time information into the reward function. 
SENTIMENT~\cite{rl_sentiment} observes that previous works do not consider the semantic information of relations in the knowledge graph, thus constructs more fine-grained types of relations by extracting comment information. 
Multi-level~\cite{rl_multi} introduces external knowledge to construct a multi-level knowledge graph and designs a Cascading Actor-Critic~\cite{haarnoja2018soft,afsar2022reinforcement} that uses a top-down strategy to search the space of the knowledge graph.
The second type of work is to combine the session information and heterogeneous information in the knowledge graph and use the item entity as the starting point. 
This type of method is represented by REKS~\cite{rs_reks}, which selects the last item in the session as the starting point for path reasoning and incorporats session information into the state representation.

Although REKS [21] incorporates sequential information to the state representation of path reasoning, the task of REKS is essentially in predicting possible paths between neighboring items, and does not take into account skip behaviors of sequential patterns present in the session when designing the reward function, and also selects only the last item in the session as the starting point for path reasoning.
However, ~\cite{rs_narm,rs_caser} point out that skip behaviors of sequential patterns are present in the session and that items with stronger correlation with the predicted goal with not necessarily appear at the end of the session.

Therefore, in order to satisfy the properties of the SR scenario, we attempt to provide a generalized explainable framework for unexplainable SR models in the form of path reasoning using a hierarchical reinforcement learning approach combined with knowledge graphs. Specifically, the reward function consists of two components: multi-target reward, which provides feedback for skip behaviors of sequential patterns that may occur in the session; and path-midpoint reward, a mechanism designed to improve the efficiency of exploration in the knowledge graph and ensure that more valuable information can be discovered. And hierarchical reinforcement learning has two levels: the session-level agent selects important items in the session as the starting point for path reasoning and then the path-level agent performs path reasoning in the knowledge graph, where important items refer to items that are highly relevant to the predicted items.

\begin{table}[]
\caption{Important notations}
\begin{tabular}{ll}
\hline
Symbol         & Description                                                                                                                                \\ \hline
$\mathcal{E}$  & The set of total entities.                                                                                                                 \\
$\mathcal{R}$  & The set of total relations.                                                                                                                \\
$\mathcal{G}$  & The knowledge graph.                                                                                                                       \\
$S_{KG^{x}}$   & Representation of entities in the knowledge graph.                                                                                         \\
$S_{se}$       & \begin{tabular}[c]{@{}l@{}}Representations of sessions obtained through \\ unexplainable SR models.\end{tabular}                           \\
$Item_{Long}$  & Relatively important item selected from session.                                                                                           \\
$Item_{Short}$ & The last item in the session.                                                                                                              \\
$S_{long}^t$   & \begin{tabular}[c]{@{}l@{}}State representation of an exploration starting\\ with $Item_{Long}$.\end{tabular}                             \\
$S_{short}^t$  & \begin{tabular}[c]{@{}l@{}}State representation of an exploration starting \\ with $Item_{Short}$.\end{tabular}                           \\
$P^{se}$       & \begin{tabular}[c]{@{}l@{}}Probability of selecting item from session \\ as $Item_{Long}$.\end{tabular}                                   \\
$P^{L}$        & \begin{tabular}[c]{@{}l@{}}Starting with $Item_{Long}$, the probability \\ of choosing the next (relation,entity).\end{tabular}            \\
$P^{S}$        & \begin{tabular}[c]{@{}l@{}}Starting with $Item_{Short}$, the probability \\ of choosing the next (relation,entity).\end{tabular}           \\
$T$            & \begin{tabular}[c]{@{}l@{}}Select T consecutive prediction target items as \\ the target items for path reasoning.\end{tabular} \\
$L^{Ce}$       & The cross-entropy loss function.                                                                                                           \\
$L^{Path}$     & Loss function for Path-level agent.                                                                                                        \\
$L^{Se}$       & Loss function for Session-level agent.                                                                                                     \\
$G_t$          & The discounted cumulative rewards at time step t.                                                                                          \\
$K$            & The length of recommendation list.                                                                                                         \\
$|A^{path}|$   & Size of the action space of the Path-level agent.                                                                                          \\ \hline
\end{tabular}
\label{table0}
\end{table}

\begin{table*}[h]
\centering
\caption{Number of the relations in the Amazon datasets.}
\begin{tabular}{llrrr}
     \hline
     Relation & Description & Beauty & Cellphones & Baby\\
     \hline
     purchase &  $user\xrightarrow{\text{purchase}} product$ & 163,678&137,832 &142,980\\
     produced\_by & $product \xrightarrow{\text{produced\_by}} brand$ &19,356 &10,222 &9,596 \\
     belong\_to & $product \xrightarrow{\text{belong\_to}} category$& 95,832 &63,178  &13,720\\
     image\_sim & $product \xrightarrow{\text{image\_sim}} image\_feature$ & 414,678 & 330,754 &359,484\\
     title\_sim & $product \xrightarrow{\text{title\_sim}} title\_feature$ &  1,567,156& 1,989,166 &967,518\\
     also\_bought & $product \xrightarrow{\text{also\_bought}} product$& 279,686 &438,812 &313,738\\
     also\_viewed & $product \xrightarrow{\text{also\_viewed}} product$& 115,218 & 17,402 &121,936\\
     viewed\_bought & $product \xrightarrow{\text{viewed\_bought}} product$&179,572& 5,190&94,698\\
     bought\_together & $product \xrightarrow{\text{bought\_together}} product$& 17,396 & 16,078 &10,154\\
     also\_bought\_diff & $product \xrightarrow{\text{also\_bought\_diff}} related\_product$& 830,874  & 676,432 &523,780\\
     also\_viewed\_diff & $product \xrightarrow{\text{also\_viewed\_diff}} related\_product$& 554,444 & 61,152 &378,040\\
     bought\_viewed\_diff& $product \xrightarrow{\text{bought\_viewed\_diff}} related\_product$& 441,260  &7,394 &128,840\\
     bought\_together\_diff& $product \xrightarrow{\text{bought\_together\_diff}} related\_product$& 11,072 &8,018 &8,956\\
     co\_occur& $product \xrightarrow{\text{co\_occur}} product$& 23,374 &  17,614 &26,281\\
     \hline
\end{tabular}
\label{table1}
\end{table*}

\begin{table}[h]
\centering
\caption{Statistics of the relations on the Douban-movie.}
\resizebox{0.95\linewidth}{!}{
\begin{tabular}{lll}
     \hline
     Relation & Description & Relation number\\
     \hline
     belong\_to & $movie\xrightarrow{\text{belong\_to}} genre$ & 41,802 \\
     directed\_by & $movie\xrightarrow{\text{directed\_by}} director$& 17,074\\
     acted\_by &$movie\xrightarrow{\text{acted\_by}} actor$ & 95,522\\
     described\_as &$movie\xrightarrow{\text{described\_as}} tag$ & 148,166 \\
     produced\_by &$movie\xrightarrow{\text{produced\_by}} region$ & 23,500\\
     \hline
\end{tabular}
}
\label{table2}
\end{table}

\section{Preliminary}

\textbf{Problem Formulation} \quad Let $\mathcal{V}$ denote the set of items and $[v_1,v_2, ...,v_n]$ denotes session history, where $v_i\in \mathcal{V} (1\leq i\leq n)$ and n denotes the length of the session. The SR model relies on the items present in the session to understand the user's interests and preferences and then make accurate predictions. The model generates a prediction score $\mathbf{y}$ for the items in the candidate pool, where $\mathbf{y}=[y_1,y_2,...,y_m]$ and m denotes the size of the candidate pool, and then selects the top k items sorted by score $y_j (1\leq j\leq m)$ as recommendations.

In our task, based on the SR model and the pre-trained knowledge graph representation, the session-level agent selects the starting point for path reasoning from the session, and the path-level agent reasons about explainable paths from the knowledge graph. We design this hierarchical reinforcement learning framework to provide explainable paths for SR models while improving recommendation accuracy.

\noindent \textbf{Knowledge Graphs} \quad Let $\mathcal{E}$ and $\mathcal{R}$ represent the entity set and relation set respectively. A knowledge graph $\mathcal{G}$ is defined as $\mathcal{G}=\{(e_h,r,e_t)|e_h,e_t \in \mathcal{E} ,r \in \mathcal{R}\}$, where each triplet $(e_h,r,e_t)$ represents a relation r from head entity $e_h$ to tail $e_t$. Our work builds a knowledge graph in conjunction with recommendation scenarios, containing multiple types of entities such as item, brand, category, etc., and also multiple types of relations such as purchase, belong\_to, produced\_by, also\_viewed, etc. A part of the relations and entities in the knowledge graph comes from the interaction between the user and the item. For example,$user_1 \xrightarrow{\text{purchase}} item_1 $ indicates that $user_1$ has purchased $item_1$. Another part of the relations and entities represent some basic attributes of the item. For example, $ item_1\xrightarrow{\text{image\_sim}} pearl $ indicates that $item_1$ has a pearl element in the image. 

\noindent \textbf{Hierarchical Reinforcement Learning}  Let $\mathcal{S}$, $\mathcal{A}$, and $\mathcal{R}$ represent the state space, action space, and reward function respectively. The hierarchical reinforcement learning (HRL) problem can be formulated as a series of Markov decision processes (MDPs) $\{MDP_1, MDP_2, ...\}$, where each markov decision process corresponds to a level or task in the hierarchy. In our task scenario, $MDP_1$ represents the session-level agent selecting the appropriate item from the session as the starting point for path reasoning, and $MDP_2$ represents the path-level agent performing path exploration in the knowledge graph.

\noindent \textbf{Explainable Paths} \quad Explainable paths contain information from both session and knowledge graph. For a session: $[v_1,v_2,v_3]$, $v_i\in \mathcal{V} (1\leq i\leq 3)$, the session-level agent chooses $v_2$ as the starting point of path reasoning, $v_2$ corresponds to the entity in the knowledge graph $\mathcal{G}$ as $item_2$, and the path-level agent starts from $item_2$ and eventually derives multiple explainable paths, e.g., $ [v_1,v_2,v_3] \xrightarrow{\text{session-level agent}} v_2  \xrightarrow{\mathcal{G}} item_2 \xrightarrow{\text{belong\_to}} category_1 \xleftarrow{\text{belong\_to}} item_4$.

Table~\ref{table0} summarizes the main symbols.

\graphicspath{{pdfs/}}
\begin{figure}[h]
    \includegraphics[width=8cm]{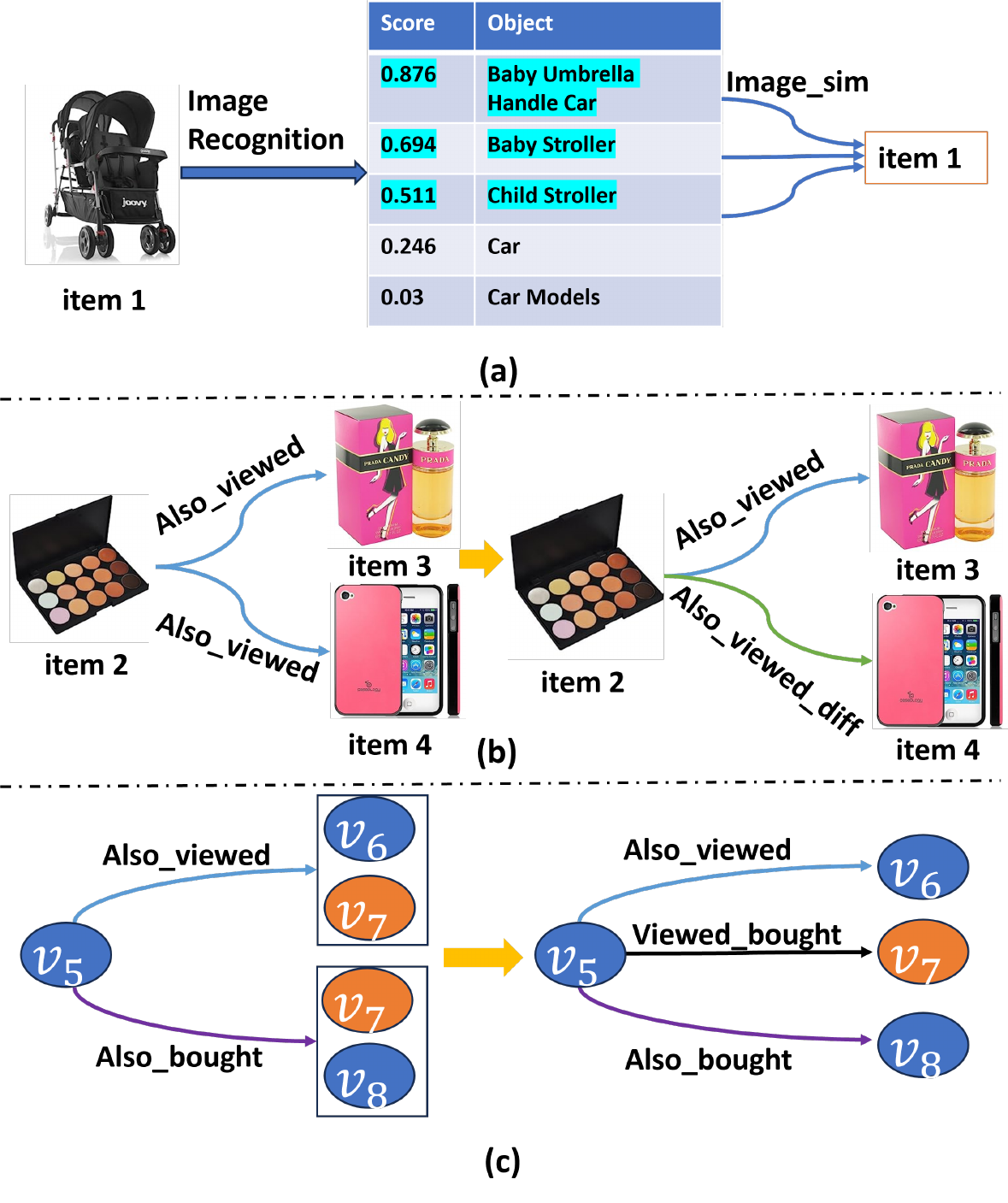}
    %1202_build_kg.pdf
    \caption{Details of knowledge graph construction: (a) extract image features; (b) split relations based on different product domains; (c) merge duplicate entities. }
    \label{fig_build_kg}
\end{figure}

\section{Knowledge Graph Construction}
As shown in Fig.~\ref{fig_build_kg}, unlike the previous path reasoning method of constructing knowledge graphs~\cite{rl_pgpr,rs_reks}, we improve the completeness and clarity of the constructed e-commerce knowledge graph in the following three ways: (a) supplementing the feature information in the product images into the knowledge graph; (b) dividing the types of relations according to the entities in different product domains; (c) merging duplicate entities.

\textbf{(a) Extract Image Feature} \quad As shown in Fig.~\ref{fig_build_kg} (a), we extract the top-5 feature information in each image through the image recognition API provided by Baidu Intelligent Cloud\footnote{\textsuperscript{https://cloud.baidu.com/product/imagerecognition}}, and select the feature information with confidence score greater than $50\%$ to be added to the knowledge graph, which connects to the items through $image\_sim$ relation. This part of information can supplement the missing feature information of the items, so that items with the same feature information can be connected, increasing the diversity of explainable paths.

\textbf{(b) Split Relations} \quad As shown in Fig.~\ref{fig_build_kg} (b), we observe that there are different types in the items connected through $also\_viwed$ or $also\_bought$, e.g., $item\enspace2$ and $item\enspace3$ belong to the cosmetic category from the Amazon-beauty dataset, while $item\enspace4$ is the cellphone case from the Amazon-cellphones dataset. We divide the original $also\_viewed$ and $also\_bought$ relations into $also\_viewed$ and $also\_viewed\_diff$, and $also\_bought$ and $also\_bought\_diff$, depending on whether the connected items belong to the same product domain, so that the relations in the knowledge graph can represent more specific meanings.

\textbf{(c) Merge Duplicate Entities.} \quad  Taking the Amazon-beauty dataset as an example, as shown in Fig.~\ref{fig_build_kg} (c), we find that the knowledge graph built up following the traditional approach has $51.43\%$ of the entities connected via $also\_viewed$ also appearing in the entities connected via $also\_bought$, and similarly, $39.19\%$ of the nodes connected via $also\_bought$ also appearing in the entities connected via $also\_viewed$. Thus, we take entities like $v_7$ that are connected by both $also\_viewed$ and $also\_bought$ and connect them via $viewed\_bought$. This approach reduces the number of connected relations by about 300,000 in the knowledge graph and improves the efficiency of path reasoning.

In the e-commerce scenario, we classify entities into seven categories: user, product, brand, category, image\_feature, title\_feature, related product, where related product indicates that the products do not belong to the same broad category, such as Amazon-Beauty and Amazon-Baby. And fourteen different types of relations.
For the movie recommendation scenario, we classify entities into six categories: movie, genre, director, actor, tag, and region. And five different types of relations. The detailed data is summarized in  Table~\ref{table1}, Table~\ref{table2}, Table~\ref{table3} and Table~\ref{table4}.

\section{Proposed Frameworks}
\begin{figure*}[!t]
    \centering
    \includegraphics[width=16cm]{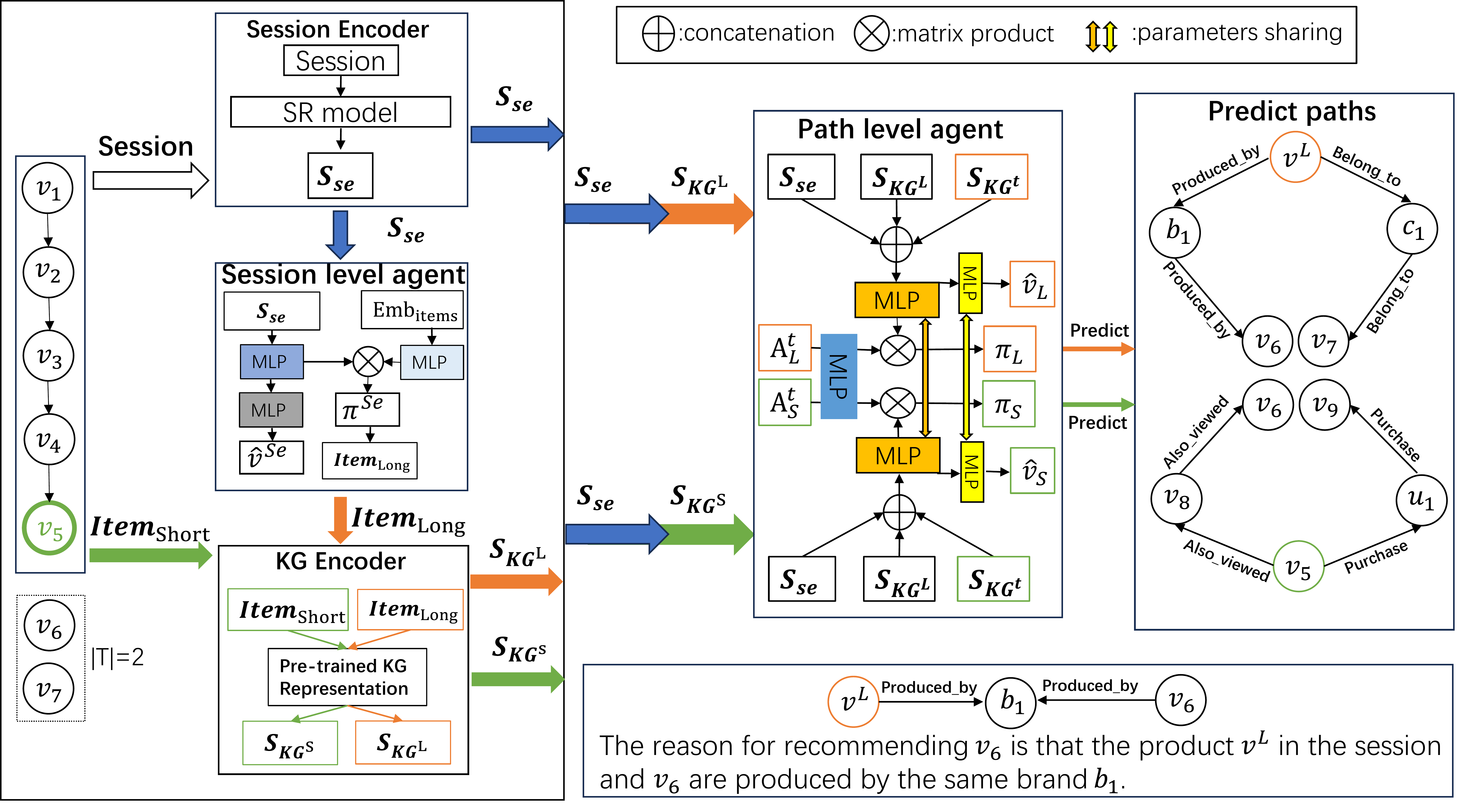}
    \caption{The overview of the proposed framework.}
    \label{framework}
\end{figure*}

The overview of the proposed framework PR4SR is illustrated in Fig.~\ref{framework}. PR4SR first utilizes the Session Encoder to encode sequence information as the state information of the session-level agent, which selects the appropriate item from the session history as the starting point for path reasoning, labeled as $Item_{Long}$. Subsequently, the last item in the session is selected as the other starting point for path reasoning, namely $Item_{Short}$. Next, the path-level agent integrates the sequence information encoded by the Session Encoder as well as the node information in the knowledge graph encoded by the KG encoder. The path-level agent explores the knowledge graph using $Item_{Long}$ and $Item_{Short}$ as starting points to determine the final prediction paths, respectively. The item located at the path's endpoint serves as the predicted target, representing the sequence of items that the user is expected to interact with over the next T steps.
In the rest of this section, we elaborate on the details of the framework.

\graphicspath{{pdfs/}}

\subsection{Mixed state encoder}
In order to capture the short-term interaction features between items, as well as the sequential information features in a session, we use a knowledge graph-level state encoder and a session-level encoder.

\noindent \textbf{Knowledge Graph Level State Encoder}\quad We obtain the entity representation through a graph-based pre-training approach (termed as a static Knowledge Graph Level State Encoder (KGLSE))~\cite{ml_transe}.

\begin{equation}
    S_{KG^{x}}=\mathrm{KGLSE}(e_{x})
    \label{eq:kg}
\end{equation}
\noindent where $e_x$ is a certain entity in the knowledge graph.

\noindent \textbf{Session-Level State Encoder}\quad Referring to the work of REKS~\cite{rs_reks}, we use the SR model as a Session Level State Encoder (SLSE).
\begin{equation}
    S_{se}=\mathrm{SLSE}(session)
    \label{eq:session}
\end{equation}
where $session=\{v_1,v_2,... ,v_n\}$, $v_i\in \mathcal{V} (1\leq i\leq n)$.
\subsection{Hierarchical Reinforcement Learning}
The key to our task is to provide an explainable recommendation process for SR, which requires selecting the important items in the session and performing path reasoning in the knowledge graph. However, the action space for selecting important items is the items within the session, whereas the action space for path reasoning encompasses the combinations of relations and entities within the knowledge graph. Therefore a single agent structure cannot solve our problem and we design the session-level agent and path-level agent.

In the process of path reasoning, the session-level agent selects the essential items from the session as the starting point of path reasoning. Considering the average length of the session, we choose an item from them as the starting point of the path, denoted as $Item_{Long}$. Meanwhile, the last item in the session is the most recent interaction behavior~\cite{rs_stamp,rs_reks}, and we also choose it as the starting point for path reasoning, denoted as $Item_{Short}$.

$Item_{Long}$ and $Item_{Short}$ are passed through Eq.\ref{eq:kg} to get the state information $S_{KG^{L}}$ and $S_{KG^{S}}$. The path-level agent performs path reasoning to explain the recommendation process starting from $Item_{Long}$ and $Item_{Short}$ respectively.
\subsection{Definition of the MDP for PR4SR}
\textbf{State.} For the session-level agent, the initial state is represented as $S_{se}$ by Eq.~\ref{eq:session}. For the path-level agent, the initial state is obtained by Eq.~\ref{eq:state_long} and Eq.~\ref{eq:state_short}.

\begin{equation}
    S_{long}^t=S_{se} \oplus S_{KG^{L}} \oplus S_{KG^{t}}    
    \label{eq:state_long}
\end{equation}
\begin{equation}
    S_{short}^t=S_{se}\oplus S_{KG^{S}} \oplus S_{KG^{t}}
    \label{eq:state_short}
\end{equation}
where $\oplus$ represents the aggregation of vectors, and $S_{KG^{t}}$ denotes the entity representation of the current node in the path at time t.

\textbf{Action.} For session-level agent, an action space $A_{se}=\{v_i|v_i\in session\}$, which denotes the selection of an item from the current session as starting point for path reasoning. For path-level agent, at time t, an action space $A^{t}_{path}=\{(r_{t+1},e_{t+1})|(e_t,r_{t+1},e_{t+1})\in \mathcal{G}\}$, which represents the selection of a  pair $(r_{t+1},e_{t+1})$ to be used at the next step.
\begin{equation}
    P^{se}=\mathrm{softmax}((W_1A_{se})\otimes(W_2S_{se}))
\end{equation}
\begin{equation}
P^{L}=\mathrm{softmax}((W_3A^t_{L})\otimes(W_4S_{long}^t))
\end{equation}
\begin{equation}
P^{S}=\mathrm{softmax}((W_3A^t_{S})\otimes(W_4S_{short}^t))
\end{equation}
where $W_1$, $W_2$, $W_3$ and $W_4$ are trainable parameter matrices and $\otimes$ is for matrix product. $A^t_{L}$ and $A^t_{S}$ denote the set of actions starting with $Item_{Long}$ and $Item_{Short}$ respectively.
It is noteworthy that $P^{L}$ and $P^{S}$ share the same parameters.

% \textbf{Transition Probability.} For the session-level agent, different sessions are handled at different times and there is no direct transfer process between them. For path-level agent, state $s_t$ is transferred to the next state $s_{t+1}$ by the action $a_t = (r_{t+1}, e_{t+1})$.
% \begin{equation}
%     P[s_{t+1}|s_t,a_t=(r_{t+1},e_{t+1})]=1
% \end{equation}

\textbf{Transition Function.} For the session-level agent, the transition function is expressed as $p[s^{se}_{t+1}|s^{se}_t,a^{se}_t=v]$, where $v\in \mathcal{V}$. For the path-level agent, the transition function is expressed as $p[s^{path}_{t+1}|s^{path}_t,a^{path}_t=(r_{t+1},e_{t+1})]$, where $(r_{t+1},e_{t+1})\in \mathcal{G}$.
\graphicspath{{pdfs/}}
\begin{figure}[h]
    \centering
    \includegraphics[width=8cm]{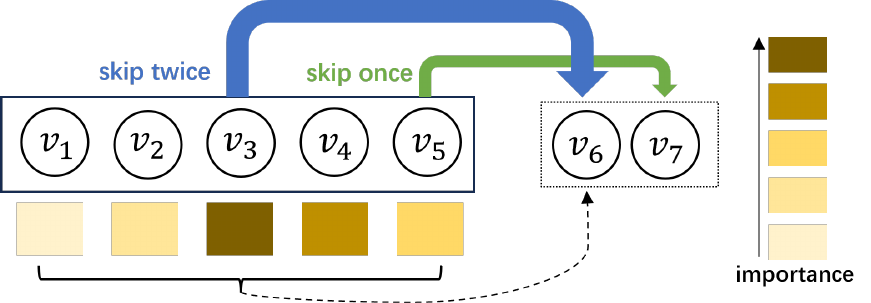}
    \caption{An example explaining the design of Multi-Target Reward.}
    \label{fig3_1}
\end{figure}

\textbf{Reward.} We designed two types of rewards.

\textbf{1) Multi-Target Reward} \quad 
Previous studies~\cite{rs_caser,rs_narm} have shown the existence of skip behaviors of sequential patterns in session history. As shown in Fig.~\ref{fig3_1},~\cite{rs_caser} found that there are strong connectivity relationships in session history such as $v_3 \rightarrow v_6$ or $v_5 \rightarrow v_7$ with multiple intermediate items in between, and ~\cite{rs_narm} also proved, from a visualization perspective, that the item that is most relevant to $v_6$ does not necessarily exist at the end of the session history.
Prior work~\cite{rs_reks} has only rewarded the $v_5 \rightarrow v_6$ explanatory path with positive feedback, but in fact, the explored $v_3 \rightarrow v_6$ or $v_5 \rightarrow v_7$ paths are also consistent with the results of the session recommendation model and are explainable paths. To ensure that valuable explainable paths explored can be given positive feedback, we set up successive T targets, which are stored in $T_{list}$ in order. 
\begin{equation}
R_{Multi} = \begin{cases}
    \mathrm{T-Index}(v_{end}) & \text{if } v_{end}\enspace in\enspace T_{list}, \\
    \log(S_{KG^{end}}*S_{KG^{tar_0}}) & \text{otherwise }.
\end{cases}
\end{equation}
where $v_{end}$ denotes the item that appears at the end of the predicted path and $\mathrm{Index}(v_{end})$ denotes the position of $v_{end}$ in $T_{list}$, $0\leq \mathrm{Index}(v_{end})<\mathrm{T}$. $tar_0$ denotes the 0th item in $T_{list}$. $S_{KG^{end}}$ and $S_{KG^{tar_0}}$ denote the state representation by Eq.~\ref{eq:kg}. The reward function indicates that: if the $v_{end}$ is in $T_{list}$, $R_{Multi}=\mathrm{T-Index}(v_{end})$; otherwise, we set the reward as the similarity between the target product $tar_0$ and the predicted $v_{end}$.
% \graphicspath{{pdfs/}}
% \begin{figure}[h]
%     \centering
%     \includegraphics[width=8cm]{pdfs/0220_reward1.pdf}
%     \caption{An example explaining the design of Multi-Target Reward.}
%     \label{fig3_1}
% \end{figure}

\graphicspath{{pdfs/}}
\begin{figure}[h!]
    \centering
    \includegraphics[width=8cm]{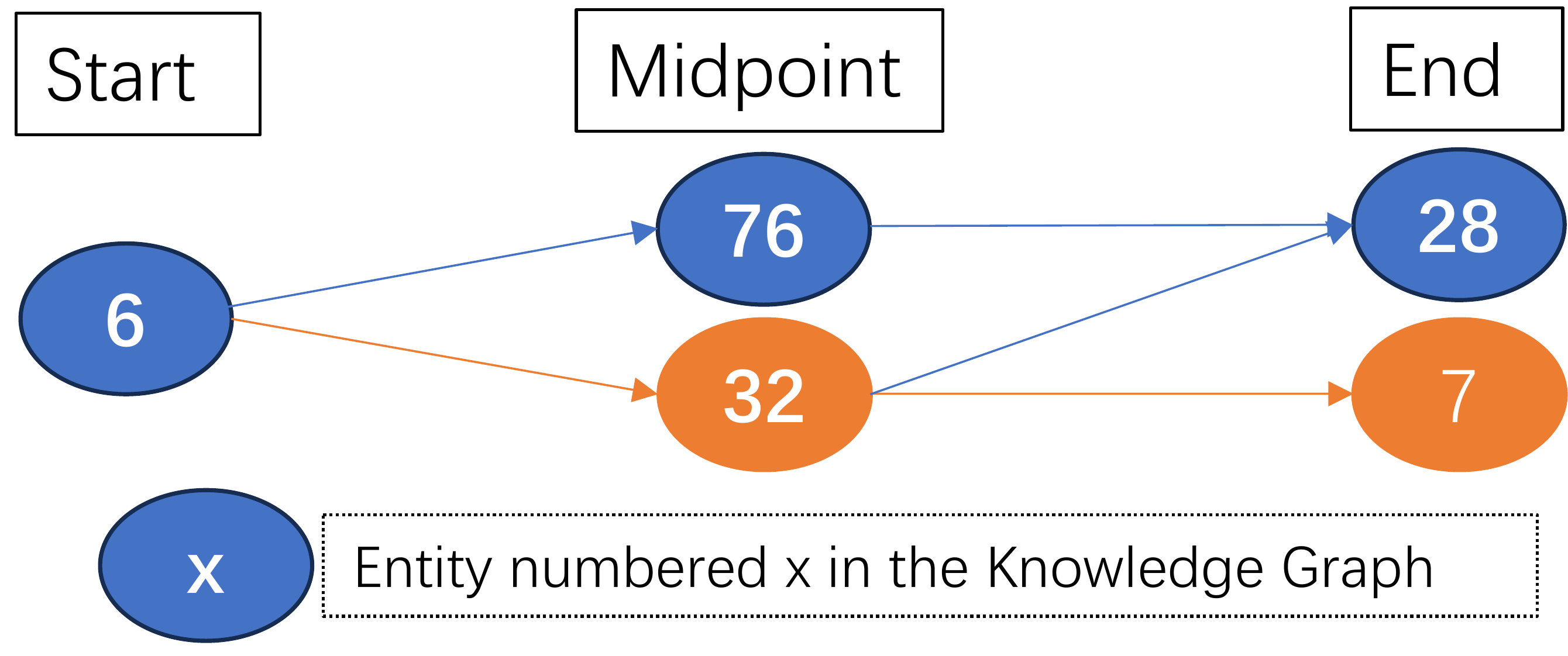}
    \caption{An example explaining the design of Path Midpoint Reward. Entity 7 is the goal point.}
    \label{fig3}
\end{figure}
\textbf{2) Path Midpoint Reward} \quad  As shown in Fig.~\ref{fig3}, we find that there is a problem in relying only on the endpoint of the path to calculate the reward.
For example, although the path ``6→32→28" and the path ``6→76→28" are both incorrect paths, the path ``6→32→28" has the node of 32, which is nearer to the goal point 7. We store the midpoint from the start point to the goal point in the dictionary $dic_{path}$, $dic_{path}[(start\enspace point,end\enspace point)]=[(r_1,e_1),...,(r_c,e_c)], (r_i,e_i) \in \mathcal{G}$, where c indicates that there are several different middle $(r_i,e_i)$.
Then we define path midpoint reward as follows:
\begin{equation}
\begin{aligned}
R_{PMid} = \max\left(0, \max_{i \in \text{range}(T)} [T - i \, | \, (r,e) \in \text{dic}^i_{path}]\right)
\end{aligned}
\end{equation}
where T denotes T successive targets, $\text{dic}^i_{path}=\text{dic}_{path}[(start\enspace point,tar_i)],\enspace tar_i \in T_{list}$. $(r,e)$ indicates the current action.
% \graphicspath{{pdfs/}}
% \begin{figure}[h!]
%     \centering
%     \includegraphics[width=8cm]{pdfs/0220_reward1.pdf}
%     \caption{An example explaining the design of Multi-Target Reward.}
%     \label{fig3_1}
% \end{figure}

% \graphicspath{{pdfs/}}
% \begin{figure}[h!]
%     \centering
%     \includegraphics[width=8cm]{pdfs/0812_midpoint.pdf}
%     \caption{An example explaining the design of Path Midpoint Reward. Entity 7 is the goal point.}
%     \label{fig3}
% \end{figure}

\textbf{Optimization.}\quad Our loss function consists of three components: session-level state encoder loss, path-level agent loss, and session-level agent loss.
\begin{equation}
    L=L^{Ce} + \alpha L^{Path} + \beta L^{Se} 
\end{equation}
where $\alpha$ and $\beta$ are hyper-parameters to balance the three parts of loss.

The path-level agent's loss function is defined as:
\begin{equation}
    G_t=\sum_{i=t+1}^{M}\gamma^{i-t-1}R_i 
\end{equation}

\begin{equation}
    L_\pi^{Path}(\theta_\pi^{path})=E_\pi[(\hat{v}^{path}(s_t)-G_t)\ln\pi(a_t|s_t,\theta_\pi^{path})]
\end{equation}
where $G_t$ represents the discounted cumulative rewards at time step $t$, M denotes the number of iteration steps, $\gamma$ indicates the discount factor, $R_{i}= R_{PMid}\enspace or \enspace R_{Multi}$. $\hat{v}^{path}(s_t)$ represents the estimate of the state value function and $\theta_\pi^{path}$ denotes the set of path level agent parameters.

The loss function of the session-level agent is defined as:
\begin{equation}
    L_\pi^{Se}(\theta_\pi^{se})=E_\pi[(\hat{v}^{se}(s)-G_{M-1})\ln\pi(a|s,\theta_\pi^{se})]
\end{equation}
where $\theta_\pi^{se}$ denotes the set of session level agent parameters and $\hat{v}^{se}(s)$ represents the estimate of the state value function.

The loss function for the session-level state encoder is defined as:
\begin{equation}
    L^{Ce}=-\sum(y_j\log(\hat{y}_j)+(1-y_j)\log(1-\hat{y}_j))
\end{equation}
where $y_j$ is the ground-truth score of item $v_j$ and $\hat{y}_j$ is the predicted score.

Algorithm 1 summarizes the process of training our model. After training the model, a probabilistic beam search algorithm~\cite{rl_adac} is used to derive the process of recommending items and provide explainable paths.

\begin{algorithm}[tb]
\caption{Overview of the PR4SR Framwork}
\label{alg:algorithm}
\begin{algorithmic}[1] %[1] enables line numbers
\STATE Pre-training knowledge graph
\STATE Initialize model parameters $\theta_\pi$ and $\theta_{\hat v}$ of each agent and parameters of the session-level state encoder
\FOR{each session}
\STATE Generate $S_{se}$ by Eq.~\ref{eq:session}
\STATE Select the $Item_{Long}$ as $e_0$ according to $\pi^{se}$ 
\STATE done=True,\enspace t=0
\WHILE{done}
\STATE Generate $S_{long}^t$ by Eq.~\ref{eq:kg} and action Space $A_L^t$
\STATE Select an action $(r_{t+1},e_{t+1})$ according to $\pi_{L}$
\STATE Get reward $R_{t}= R_{PMid}\enspace or \enspace R_{Multi}$.
\STATE Store $(e_t,R_t,(r_{t+1},e_{t+1}))$ into path
\STATE t=t+1
\IF{t==2}
\STATE done=False
\ENDIF
\ENDWHILE 
\STATE Select the $Item_{Short}$ as $e_0$
\STATE done=True,\enspace t=0
\WHILE{done}
\STATE Generate $S_{short}^t$ by Eq.~\ref{eq:kg} and action Space $A_S^t$
\STATE Select an action $(r_{t+1},e_{t+1})$ according to $\pi_{S}$
\STATE Get reward $R_{t}= R_{PMid}\enspace or \enspace R_{Multi}$.
\STATE Store $(e_t,R_t,(r_{t+1},e_{t+1}))$ into path
\STATE t=t+1 
\IF{t==2}
\STATE done=False
\ENDIF
\ENDWHILE
\STATE Compute $L^{Se}(\theta_{\hat{v}}^{se},\theta_{\pi}^{se}),L^{Path}(\theta_{\hat{v}}^{path},\theta_{\pi}^{path})$\enspace and $L^{Ce}$
\STATE Update parameters with Adam Gradient Descent
\ENDFOR
\end{algorithmic}
\end{algorithm}

\section{Experiments}
We conduct experiments on four real-world datasets to study RP4SR. We aim to answer the following research questions:

\textbf{RQ1}:\enspace How does PR4SR improve the performance of the session-based recommendation methods and the well-performed explainable framework?

\textbf{RQ2}:\enspace How do different components of PR4SR affect the model performance?

\textbf{RQ3}:\enspace How do different hyper-parameter settings affect the performance of PR4SR?

\textbf{RQ4}:\enspace How well does PR4SR perform on the explanation task compared to the existing explainable session-based framework?
\subsection{Experimental Setup}
\textbf{Datasets.}\quad We consider the four real-world datasets, including three datasets Beauty, Cellphones, Baby from the Amazon e-commerce platform~\cite{da_amazon}, and douban-movie from Douban\footnote{\textsuperscript{https://movie.douban.com/}}. We extract multiple types of relations and entities from the dataset. We consider a user's interactions that occurred in one day as a session, and filter out items with less than 5 interactions and sessions with lengths smaller than 2. We randomly sampled 75\% of the session data as the training set, 10\% of the data as the validation set, and 15\% of the data as the test set. Detailed data are described in  Table~\ref{table3}, Table~\ref{table4} and Table~\ref{table5}.

% data
\begin{table}[h]
\centering
\caption{Number of the entities in the Amazon datasets.}
\begin{tabular}{rrrr}
     \hline
     Entity & Beauty & Cellphones & Baby \\
     \hline
     user & 15,438 & 17,933 & 13,655\\ 
     product & 11,673 & 9,805 & 6,860 \\
     brand & 2,008 & 904 & 716  \\
     category & 238 & 107 & 1 \\
     image\_feature & 2,677 &1,256 & 2,136 \\
     title\_feature & 11,647 &10,202 & 6,270 \\
     related\_product & 160,281 & 96,674 & 68,168 \\
     \hline
\end{tabular}
\label{table3}

\end{table}

\begin{table}[h]
\centering
\caption{Number of Entities in the Douban-movie dataset.}
\resizebox{0.96\linewidth}{!}{
\begin{tabular}{ccccccc}
     \hline
     entity&moive&genre&director&actor&tag&region\\
     \hline
     numbers&19,114&40&6,975&34,497&16,761&373\\
     \hline
\end{tabular}
}
\label{table4}
\end{table}

\begin{table}[h]
\centering
\caption{Statistics of Amazon and Douban-movie datasets.}
\resizebox{\linewidth}{!}{
    \begin{tabular}{lrrrr}
    \hline
    Dataset        & Beauty    & Cellphones & Baby      & Douban  \\ \hline
    \#entities     & 204,007   & 136,811    & 97,851     & 77,760   \\
    \#relations    & 4,566,296 & 3,779,244  & 3,099,721 & 326,064 \\
    \#sessions     & 198,502   & 194,439    & 160,792   & 216,992 \\
    average length & 2.96      & 2.75       & 2.014     & 2.17    \\ \hline
    \end{tabular}
}
\label{table5}
\end{table}

% \usepackage{multirow}
\begin{table}[]
\centering
\caption{Hyper-parameter Settings.}
\resizebox{\linewidth}{!}{
    \begin{tabular}{lllll}
    \hline
    Methasod                                            & Dataset    & learning rate & $\alpha$     & $\beta$      \\ \hline
    \multicolumn{1}{c}{\multirow{4}{*}{PR4SR\_GRU4Rec}} & Beauty     & 0.0001        & 0.005 & 0.005  \\
    \multicolumn{1}{c}{}                                & Cellphones & 0.0001        & 0.05  & 0.01 \\
    \multicolumn{1}{c}{}                                & Baby       & 0.00005       & 0.01  & 0.005  \\
    \multicolumn{1}{c}{}                                & Douban     & 0.001         & 0.01  & 0.005  \\ \hline
    \multirow{4}{*}{PR4SR\_NARM}                        & Beauty     & 0.0001        & 0.01  & 0.005  \\
                                                        & Cellphones & 0.0001        & 0.005 & 0.01 \\
                                                        & Baby       & 0.0005        & 0.01  & 0.005  \\
                                                        & Douban     & 0.001         & 0.005 & 0.0075 \\ \hline
    \multirow{4}{*}{PR4SR\_GCSAN}                       & Beauty     & 0.0001        & 0.01  & 0.005  \\
                                                        & Cellphones & 0.0001        & 0.005 & 0.01   \\
                                                        & Baby       & 0.0005        & 0.005 & 0.005  \\
                                                        & Douban     & 0.001         & 0.01  & 0.0075 \\ \hline
    \multirow{4}{*}{PR4SR\_SR-GNN}                      & Beauty     & 0.0001        & 0.01  & 0.005  \\
                                                        & Cellphones & 0.0001        & 0.005 & 0.01   \\
                                                        & Baby       & 0.0005        & 0.005 & 0.005  \\
                                                        & Douban     & 0.001         & 0.01  & 0.0075 \\ \hline
    \multirow{4}{*}{PR4SR\_SASRec}                      & Beauty     & 0.0001        & 0.01  & 0.005  \\
                                                        & Cellphones & 0.0001        & 0.05  & 0.01   \\
                                                        & Baby       & 0.0005        & 0.01  & 0.005  \\
                                                        & Douban     & 0.001         & 0.05  & 0.01   \\ \hline
    \end{tabular}
}
\label{hyper}
\end{table}

\begin{table*}[!h]
\centering
\caption{Overall performance comparison on the four datasets.}
\resizebox{\textwidth}{!}{
% \usepackage{multirow}
\begin{tabular}{ll|rrr|rrr|rrr|rrr|rrr|rr}
\hline
\multirow{2}{*}{Dataset} & \multirow{2}{*}{Metric} & \multicolumn{3}{c|}{GRU4Rec}   & \multicolumn{3}{c|}{NARM}      & \multicolumn{3}{c|}{GCSAN}     & \multicolumn{3}{c|}{SR-GNN}    & \multicolumn{3}{c|}{SASRec}    & REKS-Avg. & NONE-Avg. \\ \cline{3-17}
                         &                         & NONE  & REKS  & PR4SR            & NONE  & REKS  & PR4SR           & NONE  & REKS  & PR4SR          & NONE  & REKS  & PR4SR            & NONE  & REKS  & PR4SR            & Improv.   & Improv.   \\ \hline
\multirow{6}{*}{Beauty}  & HR@5                    & 8.70  & 9.96  & \textbf{10.98} & 9.48  & 11.14 & \textbf{12.57} & 7.87  & 9.19  & \textbf{9.56}  & 9.43  & 9.78  & \textbf{10.41} & 9.00  & 11.38 & \textbf{12.95} & 9.46\%    & 26.92\%   \\
                         & HR@10                   & 10.98 & 13.57 & \textbf{15.41} & 11.88 & 14.82 & \textbf{17.01} & 10.49 & 12.88 & \textbf{14.02} & 11.62 & 12.89 & \textbf{15.41} & 14.11 & 15.34 & \textbf{17.26} & 13.84\%   & 34.40\%   \\
                         & HR@20                   & 14.00 & 17.62 & \textbf{20.13} & 14.67 & 19.11 & \textbf{21.93} & 13.09 & 17.30 & \textbf{19.91} & 14.21 & 17.10 & \textbf{21.26} & 19.56 & 20.00 & \textbf{22.16} & 15.83\%   & 41.64\%   \\
                         & NDCG@5                  & 6.49  & 6.82  & \textbf{7.55}  & 7.12  & 7.82  & \textbf{8.85}  & 5.76  & 6.19  & \textbf{6.46}  & 6.92  & 6.82  & \textbf{7.08}  & 5.97  & 7.91  & \textbf{9.07}  & 9.36\%    & 21.41\%   \\
                         & NDCG@10                 & 7.23  & 7.99  & \textbf{8.98}  & 7.89  & 9.00  & \textbf{10.28} & 6.61  & 7.36  & \textbf{7.90}  & 7.63  & 7.82  & \textbf{8.70}  & 7.61  & 9.19  & \textbf{10.46} & 11.79\%   & 25.08\%   \\
                         & NDCG@20                 & 7.99  & 9.01  & \textbf{10.17} & 8.59  & 10.09 & \textbf{11.52} & 7.26  & 8.48  & \textbf{9.40}  & 8.28  & 8.89  & \textbf{10.17} & 8.99  & 10.36 & \textbf{11.77} & 13.19\%   & 28.93\%   \\ \hline
\multirow{6}{*}{Cellphones}   & HR@5                    & 7.22  & 7.06  & \textbf{9.68}  & 8.16  & 7.77  & \textbf{9.97}  & 6.85  & 8.08  & \textbf{9.16}  & 7.16  & 6.95  & \textbf{9.39}  & 6.12  & 8.11  & \textbf{10.11} & 27.70\%   & 37.27\%   \\
                         & HR@10                   & 9.62  & 10.76 & \textbf{13.72} & 10.98 & 10.94 & \textbf{14.16} & 9.48  & 11.54 & \textbf{13.09} & 9.87  & 10.21 & \textbf{13.50} & 11.07 & 11.10 & \textbf{14.24} & 26.15\%   & 35.00\%   \\
                         & HR@20                   & 12.91 & 15.70 & \textbf{18.70} & 14.30 & 15.25 & \textbf{19.62} & 12.81 & 16.12 & \textbf{18.16} & 13.23 & 13.82 & \textbf{19.03} & 17.59 & 15.26 & \textbf{20.02} & 25.86\%   & 36.29\%   \\
                         & NDCG@5                  & 5.10  & 4.51  & \textbf{6.47}  & 5.78  & 5.20  & \textbf{6.76}  & 4.66  & 5.19  & \textbf{6.17}  & 5.06  & 4.63  & \textbf{6.29}  & 3.58  & 5.43  & \textbf{6.65}  & 30.14\%   & 37.28\%   \\
                         & NDCG@10                 & 5.87  & 5.70  & \textbf{7.78}  & 6.71  & 6.21  & \textbf{8.11}  & 5.50  & 6.31  & \textbf{7.43}  & 5.95  & 5.67  & \textbf{7.62}  & 5.16  & 6.40  & \textbf{8.01}  & 28.84\%   & 34.33\%   \\
                         & NDCG@20                 & 6.70  & 6.95  & \textbf{9.04}  & 7.55  & 7.30  & \textbf{9.48}  & 6.34  & 7.47  & \textbf{8.70}  & 6.79  & 6.58  & \textbf{9.01}  & 6.80  & 7.45  & \textbf{9.39}  & 27.91\%   & 33.69\%   \\ \hline
\multirow{6}{*}{Baby}    & HR@5                    & 4.83  & 5.17  & \textbf{5.83}  & 5.15  & 5.67  & \textbf{6.05}  & 3.56  & 5.59  & \textbf{5.72}  & 3.90  & 5.37  & \textbf{5.47}  & 5.01  & 5.67  & \textbf{6.14}  & 6.34\%    & 32.30\%   \\
                         & HR@10                   & 7.41  & 7.53  & \textbf{8.47}  & 7.37  & 8.05  & \textbf{8.86}  & 5.97  & 7.97  & \textbf{8.39}  & 6.22  & 7.90  & \textbf{8.28}  & 7.22  & 8.09  & \textbf{8.91}  & 8.54\%    & 26.30\%   \\
                         & HR@20                   & 10.92 & 10.87 & \textbf{12.25} & 11.01 & 11.06 & \textbf{12.71} & 9.63  & 11.04 & \textbf{12.24} & 9.31  & 10.89 & \textbf{11.93} & 10.79 & 11.50 & \textbf{12.32} & 11.06\%   & 19.42\%   \\
                         & NDCG@5                  & 2.86  & 3.47  & \textbf{4.07}  & 3.25  & 3.94  & \textbf{4.14}  & 2.43  & 3.82  & \textbf{3.91}  & 2.44  & 3.57  & \textbf{3.72}  & 3.34  & 3.93  & \textbf{4.15}  & 6.99\%    & 41.53\%   \\
                         & NDCG@10                 & 3.69  & 4.23  & \textbf{4.91}  & 3.96  & 4.71  & \textbf{5.05}  & 3.01  & 4.59  & \textbf{4.76}  & 3.18  & 4.38  & \textbf{4.62}  & 4.04  & 4.71  & \textbf{5.05}  & 7.96\%    & 37.78\%   \\
                         & NDCG@20                 & 4.57  & 5.07  & \textbf{5.87}  & 4.89  & 5.46  & \textbf{6.02}  & 3.92  & 5.36  & \textbf{5.73}  & 3.96  & 5.14  & \textbf{5.54}  & 4.94  & 5.56  & \textbf{5.90}  & 9.36\%    & 31.39\%   \\ \hline
\multirow{6}{*}{Douban}  & HR@5                    & 3.26  & 4.69  & \textbf{4.94}  & 3.04  & 4.98  & \textbf{5.32}  & 2.80  & 5.07  & \textbf{5.07}  & 3.40  & 4.58  & \textbf{4.89}  & 4.00  & 4.87  & \textbf{5.48}  & 6.33\%    & 57.76\%   \\
                         & HR@10                   & 6.32  & 7.10  & \textbf{7.38}  & 6.78  & 7.25  & \textbf{8.28}  & 6.00  & 7.35  & \textbf{7.62}  & 6.81  & 7.04  & \textbf{7.53}  & 7.23  & 7.05  & \textbf{8.55}  & 10.00\%   & 18.96\%   \\
                         & HR@20                   & 8.97  & 9.48  & \textbf{10.68} & 10.53 & 9.17  & \textbf{11.66} & 8.33  & 8.97  & \textbf{10.56} & 10.78 & 9.46  & \textbf{11.11} & 11.17 & 8.94  & \textbf{11.80} & 21.42\%   & 13.07\%   \\
                         & NDCG@5                  & 1.81  & 2.69  & \textbf{3.01}  & 1.71  & 2.88  & \textbf{3.31}  & 1.52  & 3.22  & \textbf{3.14}  & 1.92  & 2.71  & \textbf{3.04}  & 2.36  & 2.84  & \textbf{3.30}  & 10.60\%   & 72.97\%   \\
                         & NDCG@10                 & 2.72  & 3.41  & \textbf{3.78}  & 2.84  & 3.57  & \textbf{4.26}  & 2.52  & 3.92  & \textbf{3.93}  & 2.97  & 3.46  & \textbf{3.85}  & 3.35  & 3.49  & \textbf{4.27}  & 12.83\%   & 40.46\%   \\
                         & NDCG@20                 & 3.09  & 3.60  & \textbf{4.43}  & 3.61  & 3.55  & \textbf{4.86}  & 2.76  & 3.71  & \textbf{4.47}  & 3.61  & 3.88  & \textbf{4.60}  & 4.17  & 3.73  & \textbf{4.93}  & 26.23\%   & 37.20\%   \\ \hline
\end{tabular}
}
\label{table6}
\end{table*}

\graphicspath{{pdfs/}}
\begin{figure*}[!t]
    \centering
    \includegraphics[width=18cm]{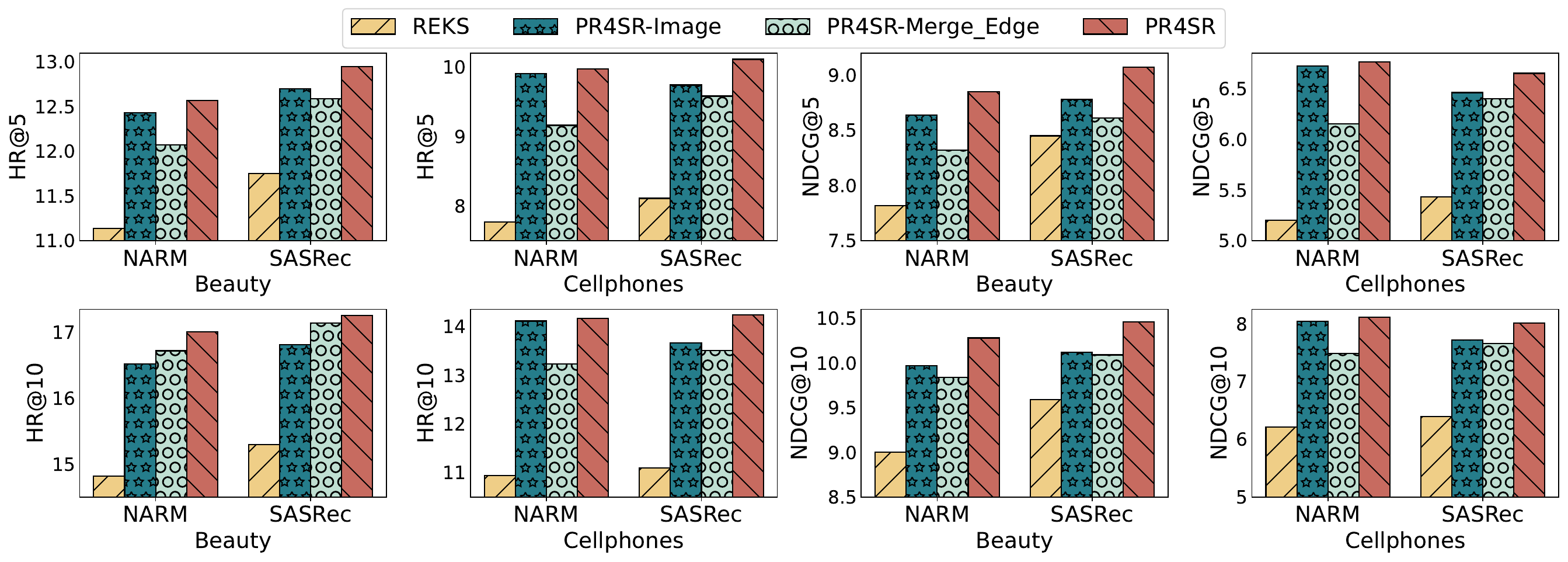}
    %1203_ablation_kg_5_10.pdf
    \caption{Ablation performance of different components in the knowledge graph.}
    \label{fig_kg}
\end{figure*}

\graphicspath{{pdfs/}}
\begin{figure*}[!t]
    \centering
    \includegraphics[width=18cm]{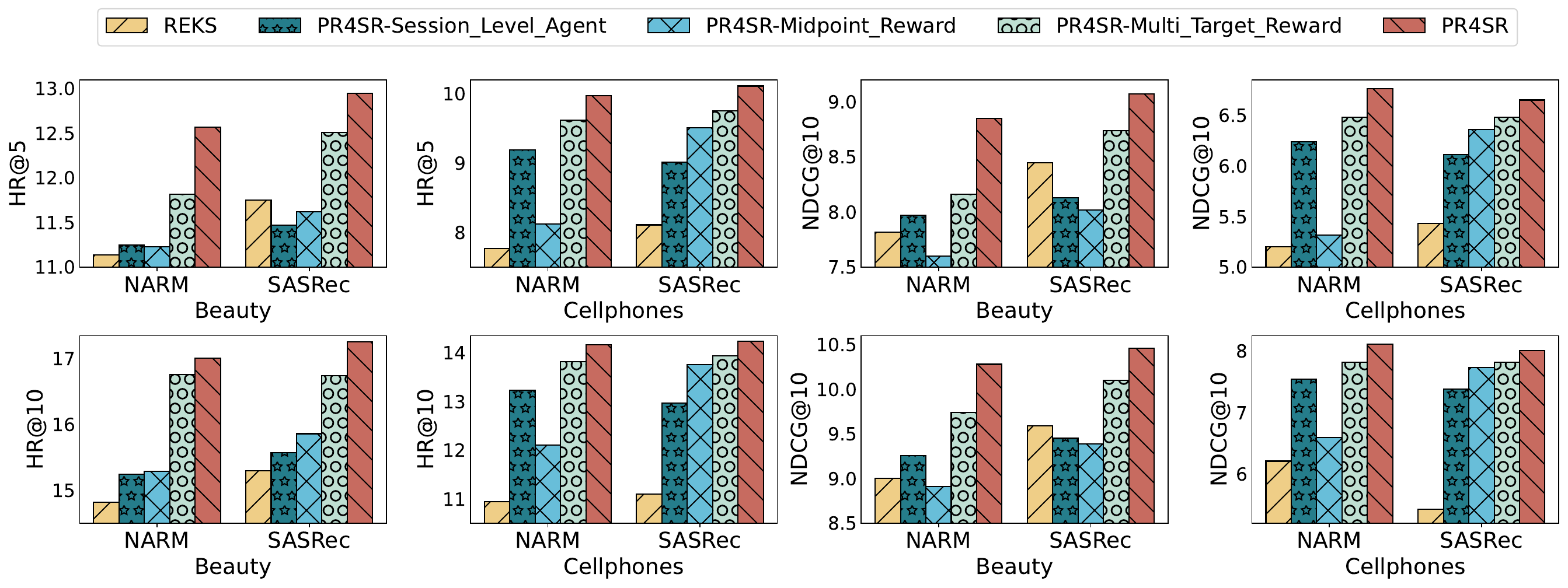}
    %1203_ablation_rl_5_10.pdf
    \caption{Ablation performance of different components in the Hierarchical Reinforcement Learning.}
    \label{fig_rl}
\end{figure*}

\textbf{Metircs.}\quad Regarding the recommendation task, We chose two metrics, HR@k (Hit Ratio), NDCG@k (Normalized Discounted Cumulative Gain), and the values of k are 5, 10, and 20. HR@k measures the proportion of recommendation lists where at least one relevant item in the top-k list is included. NDCG@k takes into account the position of the correctly recommended item in the top-k list, the higher the position, the higher the score. A larger value indicates better performance for both metrics. Regarding the explanation task, we use a questionnaire approach.

\textbf{Baselines.}\quad We apply PR4SR on five representative and widely-used SR baselines, i.e., one RNN-based method GRU4REC~\cite{rs_gru4rec}, one hybrid method that combines attention and RNN NARM~\cite{rs_narm}, two GNN-based models GCSAN~\cite{rs_gcsan} and SR-GNN~\cite{rs_srgnn}, and one attention-based method SASRec~\cite{rs_sasrec}. An explainable SR framework REKS\footnote{\textsuperscript{To the best of our knowledge, REKS is the only explainable SR framework before.}}~\cite{rs_reks}.

\textbf{Reproducibility Settings.}\quad We followed the original settings suggested by the authors~\cite{rs_reks} to train baseline model (i.e., GRU4Rec, NARM, GCSAN, SR-GNN) on Amazon dataset. For the recommendation process, the path length is set to 2, the maximum action space is set to 200, and the action dropout is set to 0.7. For the reinforcement learning design, the discount factor $\gamma$ is set to 0.99, and $W_1\in R^{400*400}, W_2\in R^{400*400}, W_3\in R^{400*400}, W_4\in R^{800*400}$. For the Amazon dataset, Beauty, Cellphones, and Baby, our model is trained for 150 epochs using Adam optimization. For the Douban dataset, our model is trained for 50 epochs using Adam optimization. The batch size is set to 256, and T is set to 5. In the test phase, the sample sizes of the two steps are set to 100 and 1 respectively. More details in Table~\ref{hyper}.

\subsection{Results and Analysis}
\textbf{RQ1: Overall Results.} 

The experimental results with all baseline methods with REKS or with our method PR4SR are illustrated in Table~\ref{table6}, where NONE denotes no use of explainable frameworks and ``REKS-Avg Imporv." denotes the average improvement in the unexplainable SR model with PR4SR relative to the unexplainable  SR model with REKS and ``NONE-Avg Imporv." denotes the average improvement in the unexplainable SR model with PR4SR relative to the unexplainable SR models. PR4SR improves the recommendation performance of unexplainable SR models with all metrics in all datasets. It also outperforms the well-performed explainable SR framework. This proves the effectiveness of PR4SR. Moreover, on the three datasets Cellphones, Baby, and Douban, when the k in top-k is smaller, PR4SR improves more obviously than unexplainable SR models. This is due to the fact that customers pay more attention to the nearest few items when shopping, further indicating that PR4SR is suitable for practical session-based scenarios.

We find that NARM and SASRec achieve the best results on multiple datasets via PR4SR compared to other SR models. And GRU4Rec also has relatively good results compared to SR-GNN and GCSAN on different datasets via PR4SR. This suggests that PR4SR is more suitable for the attention mechanism and RNN-based SR models, compared to the GNN-based SR models.

% \usepackage{multirow}

\graphicspath{{pdfs/}}
\begin{figure*}[!t]
    \centering
    \includegraphics[width=18cm]{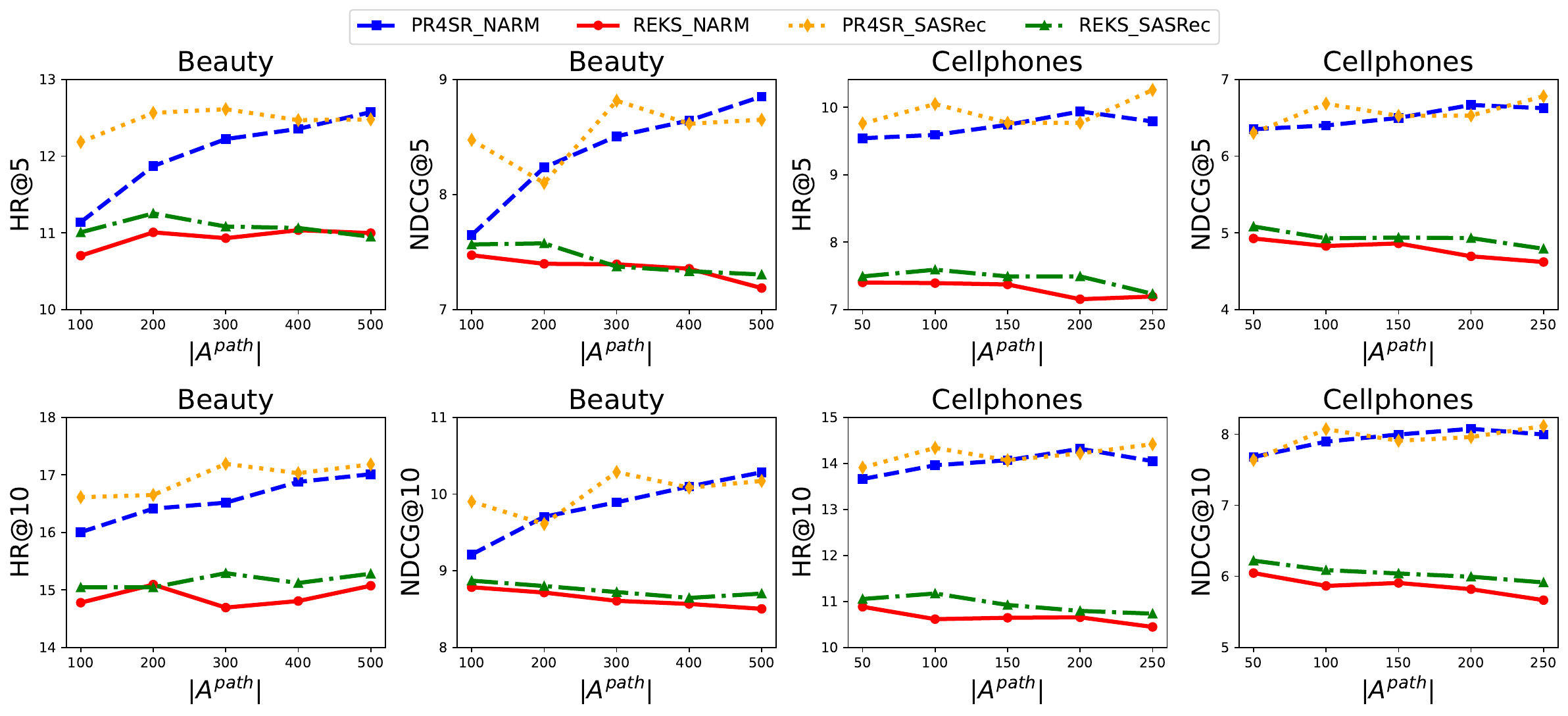}
    %pdfs/1128_action_space_5_10.pdf
    \caption{Impact of path-level agent's action space size on performance (K = 5,10).}
    \label{fig4}
\end{figure*}

% \graphicspath{{pdfs/}}
% \begin{figure*}[!t]
%     \centering
%     \includegraphics[width=18cm]{1203_hyper.pdf}
%     %1203.pdf
%     \caption{ Impact of different hyper-parameters (K = 5, 10).}
%     \label{fig5}
% \end{figure*}

\noindent\textbf{RQ2: Ablation Study.}

As shown in Fig.~\ref{fig_kg} and Fig.~\ref{fig_rl}, based on NAMR and SASRec, we design ablation experiments to test the design of innovations in the model from two aspects: knowledge graph (KG) and hierarchical reinforcement learning (HRL).

\textbf{KG}: (1) PR4SR-Image denotes the removal of image\_feature entity and image\_sim relation.  (2) PR4SR-Merge\_Edge indicates not merging the entities appearing in also\_bought and also\_viewed. From the results, we can see that all components have an impact on the improvement of the model performance, even though the addition of image information increases exploration space for the path-level agent. Although adding information about the image has little impact on the accuracy of the recommendation, image\_feature entity and image\_sim relation increase the diversity to the explanation paths. The ablation experiments after dividing also\_viewed and also\_bought according to different product domains are not considered, as the number of entities and relations in the knowledge graph does not change before and after the division, which produces a relatively small impact on the experiments, but this division method would have provided diversity of explainable paths.

\textbf{HRL}: (1) PR4SR-Session\_Level\_Agent indicates that the session-level agent is not applied, and only the last item of the session is considered to do path reasoning. (2) PR4SR-Midpoint\_Reward indicates no consideration of Path Midpoint Reward. (3) PR4SR-Multi\_Target\_Reward indicates no consideration of Multi-Target Reward, but only a single target point. The effect of removing the session-level agent on the model is relatively large, which suggests that we have selected the relatively important items in the session through the session-level agent to be the starting point of the path reasoning. The results also show that the two reward designs also help to improve the performance of the model.

\graphicspath{{pdfs/}}
\begin{figure*}[!t]
    \centering
    \includegraphics[width=18cm]{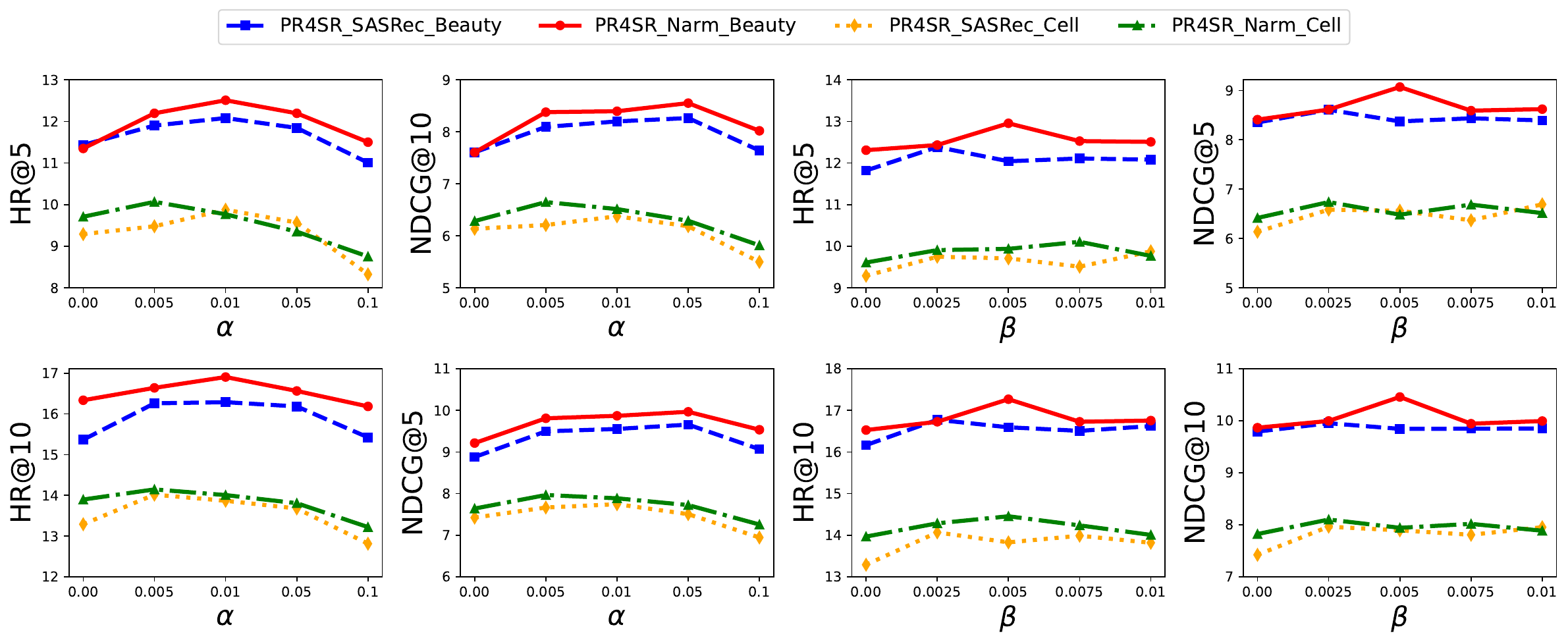}
    %1203.pdf
    \caption{ Impact of different hyper-parameters (K = 5, 10).}
    \label{fig5}
\end{figure*}
\noindent\textbf{RQ3: Sensitivity of Hyper-parameters}

\textbf{Action space size.} Since the product knowledge graph increases the number of entities and relations as new products are added, it is important to increase the RL action space while ensuring that RL can explore valuable information. From Fig. ~\ref{fig4}, it can be seen that in most cases the performance of REKS decreases or slightly improves as the action space increases, while the performance of PR4SR shows a steady increase. It further shows that PR4SR is able to adapt to changing knowledge graphs compared to REKS.

\begin{table}[!h]
\centering
\caption{The effect of the length \textbar T\textbar\ on Beauty.}
\begin{tabular}{|c|cc|cc|}
\hline
\multicolumn{1}{|c|}{\multirow{2}{*}{Setting}} & \multicolumn{2}{c|}{NARM}                   & \multicolumn{2}{c|}{SASRec}                \\ \cline{2-5} 
\multicolumn{1}{|c|}{}                          & \multicolumn{1}{l|}{HR@5} & NDCG@5          & \multicolumn{1}{l|}{HR@5} & NDCG@5         \\ \hline
T=1                                             & 11.82                   & 8.16         & 12.81                  & 8.80         \\
T=2                                             & 12.20                   & 8.53          & 12.54                   & 8.60         \\
T=3                                             & 12.04                   & 8.32           & \textbf{12.84}          & 8.84         \\
T=4                                             & 12.05                   & 8.42          & 12.75                   & \textbf{8.91} \\
T=5                                             & \textbf{12.50}          & \textbf{8.83} & 12.32                   & 8.50   \\ \hline
\end{tabular}
\label{table8}
\end{table}

\textbf{Length of successive targets.} As shown in Table~\ref{table8}, we find that PR4SR\_NARM and PR4SR\_SASRec achieve good results at T=5 and T=3, respectively. The experimental results indicate that the optimal T is greater than 2, further illustrating the effectiveness of the multi-target reward.

\textbf{Impact of $\alpha,\beta$ in loss function.} We further validate the role of $\alpha$ and $\beta$ in the loss function. Fig.~\ref{fig5} shows the results for different parameters, including $\alpha$  and $\beta$. The framework performance is  relatively stable when $\alpha$ is between 0.005 and 0.1 and $\beta$ is between 0.0025 and 0.01, which further illustrates the robustness of our framework.

\noindent \textbf{RQ4: Model Performance on Explanation Task}

\begin{figure}[h]
    \centering
    \includegraphics[width=8cm]{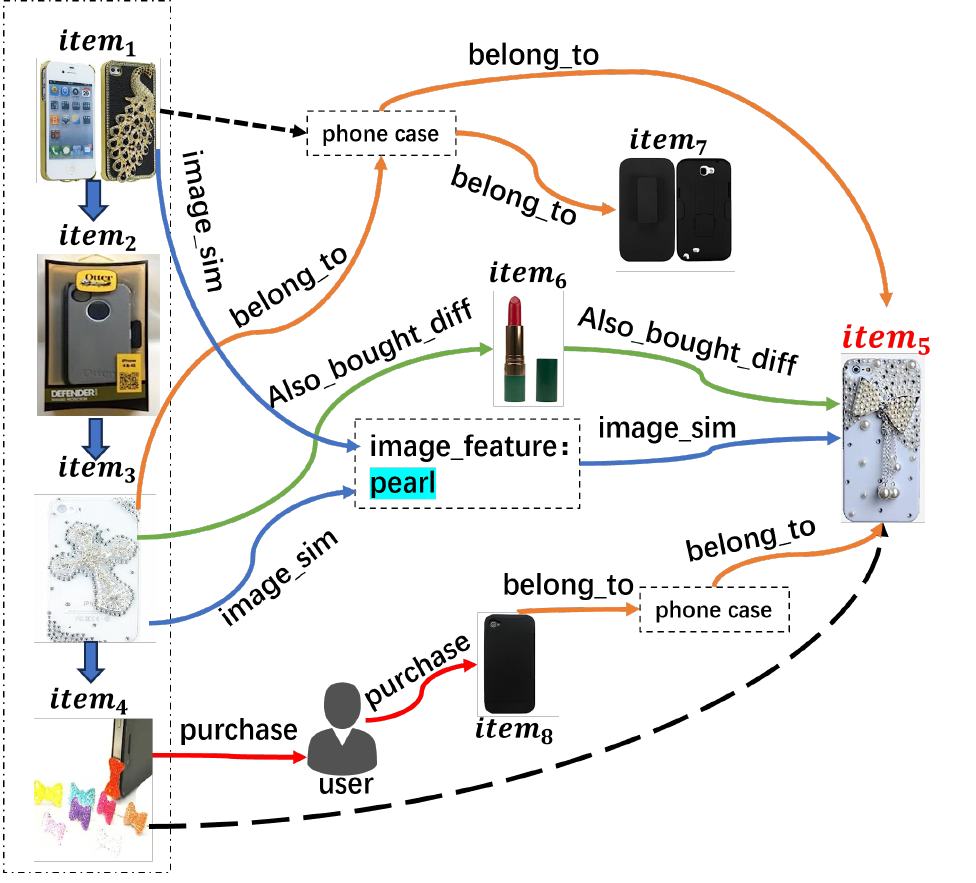}
    \caption{An example of explainable paths generated by our method. The direction of the arrow on the right side of the picture indicates the direction of path reasoning.}
    \label{fig6}
\end{figure}

\textbf{Case Study.} The case study in Fig.~\ref{fig6} demonstrates that our proposed hierarchical reinforcement learning framework can better provide interpretability for SR models as well as strategies for constructing knowledge graphs making connections between entities stronger, where the session record is ${item_1,item_2,item_3,item_4}$ and the predicted item is $item_5$.

First, our proposed model PR4SR improves the accuracy of prediction. PR4SR selects $item_4$ as the starting point of path reasoning from the session by the session-level agent, and there are multiple inference paths of length 2 between $item_3$ and the predicted entity $item_5$, e.g., $ item_3\xrightarrow{\text{belong\_to}} phone\enspace case \xleftarrow{\text{belong\_to}} item_5 $, which indicates that the reason for recommending $item_5$ is because $item_5$ and $item_3$ both have a distinct image feature-``pearl". Whereas the REKS only considered $item_4$ as the starting point for path reasoning, under the restriction of exploring paths of length 2, there does not exist a path connecting $item_4$ to the predicted entity $item_5$ and exists only  $ item_4\xleftarrow{\text{purchase}} user \xrightarrow{\text{purchase}} item_8 $.

Second, the method of constructing the knowledge graph complements the information characterizing the items and increases the diversity of explanations. As in the Fig.~\ref{fig6} ``pearl" indicates that the picture of the item contains pearls. If the knowledge graph is constructed without considering the feature information of the image, due to the lack of attribute information of $item_1$ in the dataset, there is no connecting relation between $item_1$ and the category ``phone case'', and $item_5$ cannot be inferred from $item_1$. By adding the newly added feature information of the image ``pearl", we can infer that there are two explainable paths,  $ item_1\xrightarrow{\text{image\_sim}} pearl \xleftarrow{\text{image\_sim}} item_5 $  and  $ item_3\xrightarrow{\text{image\_sim}} pearl \xleftarrow{\text{image\_sim}} item_5 $. There exists a path, $ item_3\xrightarrow{\text{also\_bought\_diff}} item_6 \xleftarrow{\text{also\_bought\_diff}} item_5 $, which states that after splitting relations based on different product domains we can determine which product class the entity belongs to by the type of the relation, increasing the explainability of the path.

\graphicspath{{pdfs/}}

\graphicspath{{pdfs/}}
\begin{figure}[h]
    \centering
    \includegraphics[width=8cm]{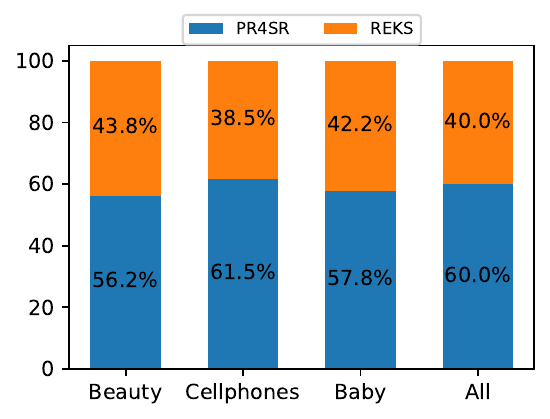}
    \caption{Percentage of user choices.}
    \label{fig7}
\end{figure}

\textbf{User Study.} 
In this study, we verify the superiority of our explanation from a human perspective. A group of 50 participants was chosen for the study. We randomly selected 30 cases from datasets Beauty, Cellphones, and Baby. Each case consisted of two different explanatory content from PR4SR and  REKS. We provided detailed data in the questionnaire, including the user’s session history, the starting point for path reasoning, and a description of the explainable path. Participants evaluated the choice of the starting point for path reasoning and the explainable paths based on their understanding of the session history.
 
The results in Fig. \ref{fig7} indicate that PR4SR outperforms REKS in terms of explainability in the case of all three datasets. This is more related to the introduction of the image\_feature and title\_feature entities, as well as image\_sim and title\_sim, which are in line with the actual buying experience of users in the shopping system.

\section{Conclusion}
In this paper, we introduce PR4SR, the first framework to utilize hierarchical reinforcement learning for providing explainable path reasoning for existing unexplainable SR models. Within this framework, the session-level agent selects key items from the session history as the path starting point and the path-level agent performs path reasoning in the knowledge graph. In particular, we motivate the learning of skip behaviors of sequential patterns in the session scenario by multi-target reward and use path-midpoint reward to improve the exploration efficiency. Meanwhile, incorporating image information into the knowledge graph enhances its completeness and the diversity of explanation paths. Extensive experiments on four datasets show that our framework outperforms current unexplainable SR models and the explainable SR framework in terms of both recommendation performance and explainability. In future research work, we will explore generic explainable frameworks for the currently existing unexplainable cross-domain recommendation models as well as constructing product knowledge graphs with more diverse explainability.

% \newpage

 

% \end{IEEEbiography}
% \begin{IEEEbiography}
% [{\includegraphics[width=1in,height=1.25in,clip,keepaspectratio]{fig1.png}}] 

% Author Introduction

% \end{IEEEbiography}



% \vfill
\begin{IEEEbiography}[{\includegraphics[width=1in,height=1.25in,clip,keepaspectratio]{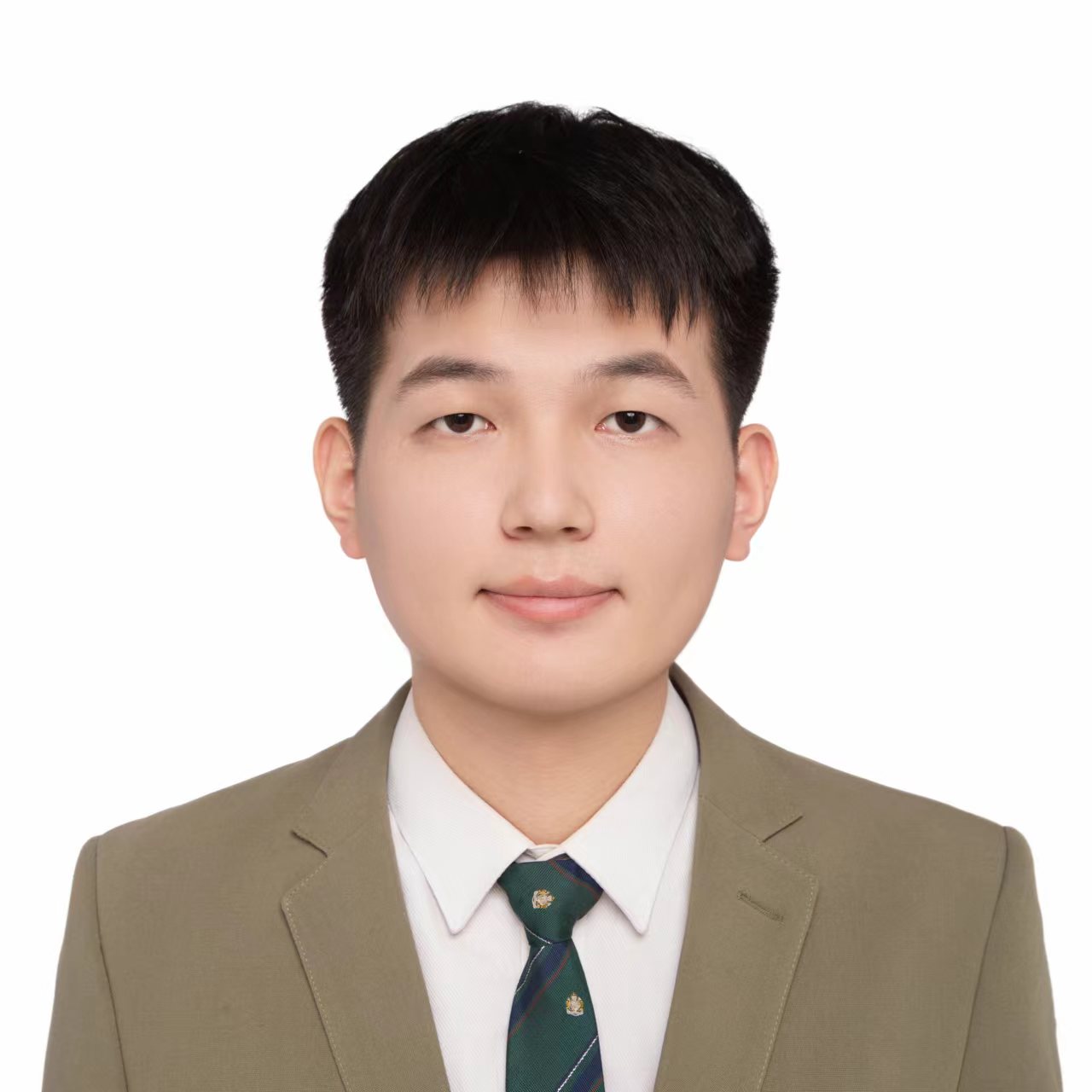}}]{Yang Cao} is currently pursuing the master's degree with the Department of Computer Science and Technology, East China Normal University, China. His research interests include recommendation system, knowledge graph and reinforcement learning.
\end{IEEEbiography}

\begin{IEEEbiography}[{\includegraphics[width=1in,height=1.25in,clip,keepaspectratio]{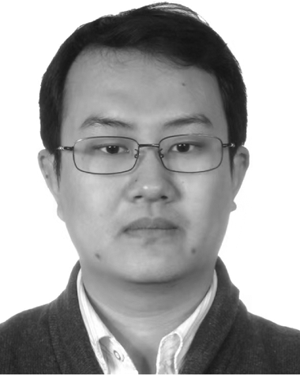}}]{Shuo Shang} is currently a professor of computer science with the University of Electronic Science and Technology of China. He was a senior scientist with the Inception Institute of Artificial Intelligence (IIAI), leading its data mining research group. His research interests include big data, data mining, and machine learning.
\end{IEEEbiography}

\begin{IEEEbiography}[{\includegraphics[width=1in,height=1.25in,clip,keepaspectratio]{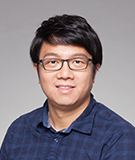}}]{Jun Wang} received the Ph.D. degree in electrical engineering from Columbia University, New York, NY, USA, in 2011. Currently, he is a professor at the School of Computer Science and Technology, East China Normal University and an adjunct faculty member of Columbia University. From 2010 to 2014, he was a research staff member at IBM T. J. Watson Research Center, Yorktown Heights, NY, USA. His research interests include machine learning, data mining, mobile intelligence, and computer vision. 
\end{IEEEbiography}

% \vspace{-16.5cm}
\begin{IEEEbiography}[{\includegraphics[width=1in,height=1.25in,clip,keepaspectratio]{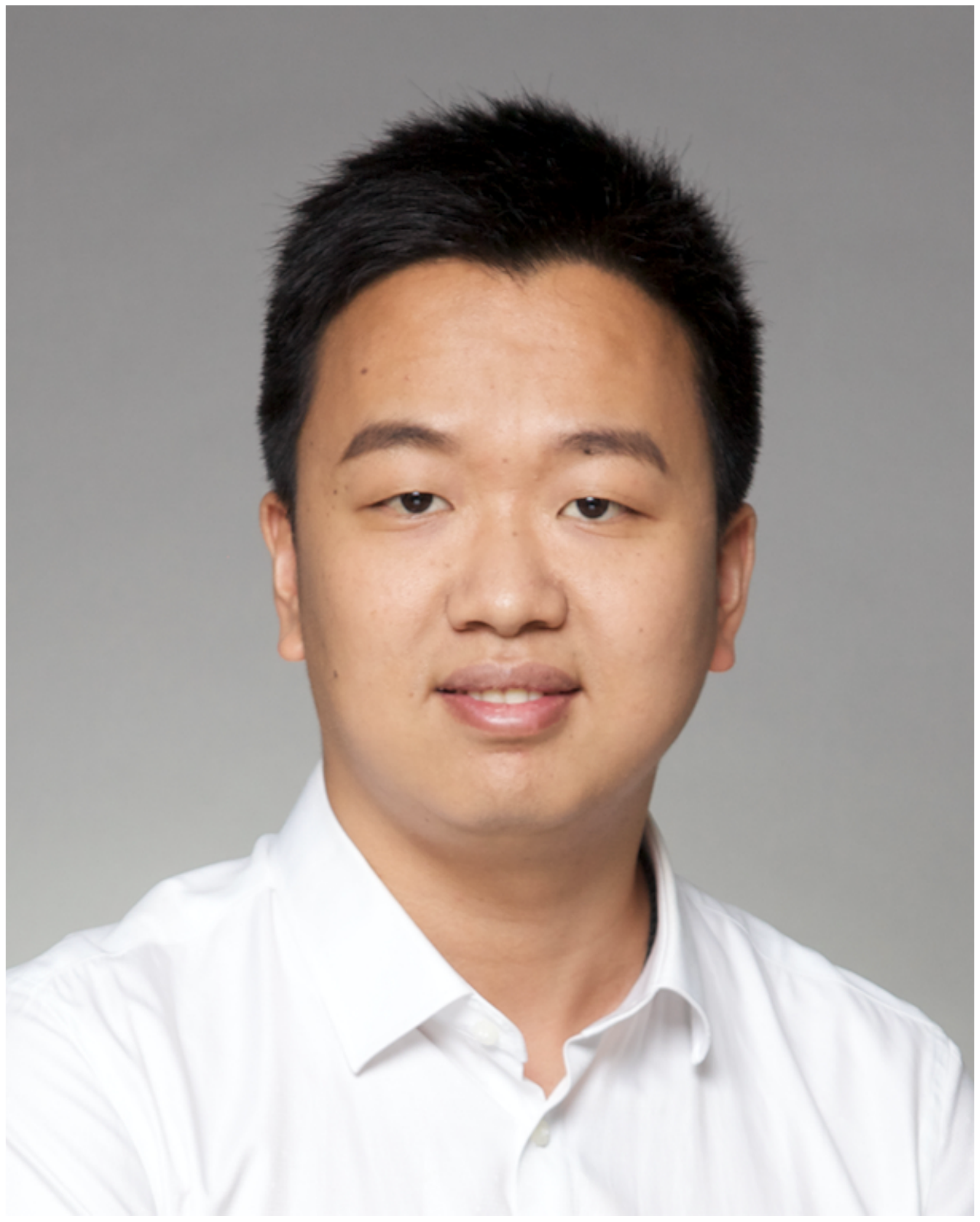}}]{Wei Zhang} received his Ph.D. degree in computer science and technology from Tsinghua university, Beijing, China, in 2016. He is currently a professor in the School of Computer Science and Technology, East China Normal University, Shanghai, China.
His research interests mainly include user data mining and machine learning applications.
He is a senior member of China Computer Federation.
\end{IEEEbiography}


\begin{thebibliography}{00}
%1
\bibitem{rs_gru4rec}
B.~Hidasi, A.~Karatzoglou, L.~Baltrunas, and D.~Tikk, ``Session-based recommendations with recurrent neural networks,'' \emph{arXiv preprint arXiv:1511.06939}, 2015.
%2
\bibitem{rs_narm}
J.~Li, P.~Ren, Z.~Chen, Z.~Ren, T.~Lian, and J.~Ma, ``Neural attentive session-based recommendation,'' in \emph{Proceedings of the 2017 ACM on Conference on Information and Knowledge Management}, 2017, pp. 1419--1428.
%3
\bibitem{rs_sasrec}
W.-C. Kang and J.~McAuley, ``Self-attentive sequential recommendation,'' in \emph{2018 IEEE international conference on data mining (ICDM)}.\hskip 1em plus 0.5em minus 0.4em\relax IEEE, 2018, pp. 197--206.
%4
\bibitem{rs_stamp}
Q.~Liu, Y.~Zeng, R.~Mokhosi, and H.~Zhang, ``Stamp: short-term attention/memory priority model for session-based recommendation,'' in \emph{Proceedings of the 24th ACM SIGKDD international conference on knowledge discovery \& data mining}, 2018, pp. 1831--1839.
%5
\bibitem{rs_gcsan}
C.~Xu, P.~Zhao, Y.~Liu, V.~S. Sheng, J.~Xu, F.~Zhuang, J.~Fang, and X.~Zhou, ``Graph contextualized self-attention network for session-based recommendation.'' in \emph{IJCAI}, vol.~19, 2019, pp. 3940--3946.
%6
\bibitem{rs_srgnn}
S.~Wu, Y.~Tang, Y.~Zhu, L.~Wang, X.~Xie, and T.~Tan, ``Session-based recommendation with graph neural networks,'' in \emph{Proceedings of the AAAI conference on artificial intelligence}, vol.~33, no.~01, 2019, pp. 346--353.

%7
\bibitem{rl_sentiment}
S.-J. Park, D.-K. Chae, H.-K. Bae, S.~Park, and S.-W. Kim, ``Reinforcement learning over sentiment-augmented knowledge graphs towards accurate and explainable recommendation,'' in \emph{Proceedings of the Fifteenth ACM International Conference on Web Search and Data Mining}, 2022, pp. 784--793.
%8
\bibitem{rl_adac}
K.~Zhao, X.~Wang, Y.~Zhang, L.~Zhao, Z.~Liu, C.~Xing, and X.~Xie, ``Leveraging demonstrations for reinforcement recommendation reasoning over knowledge graphs,'' in \emph{Proceedings of the 43rd international ACM SIGIR conference on research and development in information retrieval}, 2020, pp. 239--248.

\bibitem{fu2020fairness}
Z.~Fu, Y.~Xian, R.~Gao, J.~Zhao, Q.~Huang, Y.~Ge, S.~Xu, S.~Geng, C.~Shah, Y.~Zhang \emph{et~al.}, ``Fairness-aware explainable recommendation over knowledge graphs,'' in \emph{Proceedings of the 43rd International ACM SIGIR Conference on Research and Development in Information Retrieval}, 2020, pp. 69--78.

\bibitem{xian2020cafe}
Y.~Xian, Z.~Fu, H.~Zhao, Y.~Ge, X.~Chen, Q.~Huang, S.~Geng, Z.~Qin, G.~De~Melo, S.~Muthukrishnan \emph{et~al.}, ``Cafe: Coarse-to-fine neural symbolic reasoning for explainable recommendation,'' in \emph{Proceedings of the 29th ACM International Conference on Information \& Knowledge Management}, 2020, pp. 1645--1654.

\bibitem{lu2023user}
H.~Lu, W.~Ma, Y.~Wang, M.~Zhang, X.~Wang, Y.~Liu, T.-S. Chua, and S.~Ma, ``User perception of recommendation explanation: Are your explanations what users need?'' \emph{ACM Transactions on Information Systems}, vol.~41, no.~2, pp. 1--31, 2023.

\bibitem{liu2023social}
C.~Liu, W.~Wu, S.~Wu, L.~Yuan, R.~Ding, F.~Zhou, and Q.~Wu, ``Social-enhanced explainable recommendation with knowledge graph,'' \emph{IEEE Transactions on Knowledge and Data Engineering}, 2023.

\bibitem{lyu2022knowledge}
Z.~Lyu, Y.~Wu, J.~Lai, M.~Yang, C.~Li, and W.~Zhou, ``Knowledge enhanced graph neural networks for explainable recommendation,'' \emph{IEEE Transactions on Knowledge and Data Engineering}, vol.~35, no.~5, pp. 4954--4968, 2022.

\bibitem{wei2023rule}
Y.~Wei, X.~Qu, X.~Wang, Y.~Ma, L.~Nie, and T.-S. Chua, ``Rule-guided counterfactual explainable recommendation,'' \emph{IEEE Transactions on Knowledge and Data Engineering}, 2023.

% 9 
\bibitem{rl_multi}
X.~Wang, K.~Liu, D.~Wang, L.~Wu, Y.~Fu, and X.~Xie, ``Multi-level recommendation reasoning over knowledge graphs with reinforcement learning,'' in \emph{Proceedings of the ACM Web Conference 2022}, 2022, pp. 2098--2108.

\bibitem{lin2018multi}
X.~V. Lin, R.~Socher, and C.~Xiong, ``Multi-hop knowledge graph reasoning with reward shaping,'' \emph{arXiv preprint arXiv:1808.10568}, 2018.

\bibitem{sun2018recurrent}
Z.~Sun, J.~Yang, J.~Zhang, A.~Bozzon, L.-K. Huang, and C.~Xu, ``Recurrent knowledge graph embedding for effective recommendation,'' in \emph{Proceedings of the 12th ACM conference on recommender systems}, 2018, pp. 297--305.

\bibitem{geng2022path}
S.~Geng, Z.~Fu, J.~Tan, Y.~Ge, G.~De~Melo, and Y.~Zhang, ``Path language modeling over knowledge graphsfor explainable recommendation,'' in \emph{Proceedings of the ACM Web Conference 2022}, 2022, pp. 946--955.

% 10-13

\bibitem{ma2019jointly}
W.~Ma, M.~Zhang, Y.~Cao, W.~Jin, C.~Wang, Y.~Liu, S.~Ma, and X.~Ren, ``Jointly learning explainable rules for recommendation with knowledge graph,'' in \emph{The world wide web conference}, 2019, pp. 1210--1221.

\bibitem{wang2019explainable}
X.~Wang, D.~Wang, C.~Xu, X.~He, Y.~Cao, and T.-S. Chua, ``Explainable reasoning over knowledge graphs for recommendation,'' in \emph{Proceedings of the AAAI conference on artificial intelligence}, vol.~33, no.~01, 2019, pp. 5329--5336.

\bibitem{zhu2020knowledge}
Q.~Zhu, X.~Zhou, J.~Wu, J.~Tan, and L.~Guo, ``A knowledge-aware attentional reasoning network for recommendation,'' in \emph{Proceedings of the AAAI conference on artificial intelligence}, vol.~34, no.~04, 2020, pp. 6999--7006.
%14
\bibitem{icarte2022reward}
R.~T. Icarte, T.~Q. Klassen, R.~Valenzano, and S.~A. McIlraith, ``Reward machines: Exploiting reward function structure in reinforcement learning,'' \emph{Journal of Artificial Intelligence Research}, vol.~73, pp. 173--208, 2022.


\bibitem{rl_pgpr}
Y.~Xian, Z.~Fu, S.~Muthukrishnan, G.~De~Melo, and Y.~Zhang, ``Reinforcement knowledge graph reasoning for explainable recommendation,'' in \emph{Proceedings of the 42nd international ACM SIGIR conference on research and development in information retrieval}, 2019, pp. 285--294.

%15
\bibitem{rs_reks}
H.~Wu, H.~Fang, Z.~Sun, C.~Geng, X.~Kong, and Y.-S. Ong, ``A generic reinforced explainable framework with knowledge graph for session-based recommendation,'' in \emph{2023 IEEE 39th International Conference on Data Engineering (ICDE)}.\hskip 1em plus 0.5em minus 0.4em\relax IEEE, 2023, pp. 1260--1272.

\bibitem{rs_caser}
J.~Tang and K.~Wang, ``Personalized top-n sequential recommendation via convolutional sequence embedding,'' in \emph{Proceedings of the eleventh ACM international conference on web search and data mining}, 2018, pp. 565--573.


\bibitem{rs_mf}
Y.~Koren, R.~Bell, and C.~Volinsky, ``Matrix factorization techniques for recommender systems,'' \emph{Computer}, vol.~42, no.~8, pp. 30--37, 2009.

\bibitem{rs_mc}
B.~Sarwar, G.~Karypis, J.~Konstan, and J.~Riedl, ``Item-based collaborative filtering recommendation algorithms,'' in \emph{Proceedings of the 10th international conference on World Wide Web}, 2001, pp. 285--295.

\bibitem{rs_fpmc}
S.~Rendle, C.~Freudenthaler, and L.~Schmidt-Thieme, ``Factorizing personalized markov chains for next-basket recommendation,'' in \emph{Proceedings of the 19th international conference on World wide web}, 2010, pp. 811--820.

\bibitem{GRU}
J.~Chung, C.~Gulcehre, K.~Cho, and Y.~Bengio, ``Empirical evaluation of gated recurrent neural networks on sequence modeling,'' \emph{arXiv preprint arXiv:1412.3555}, 2014.

\bibitem{RNN}
W.~Zaremba, I.~Sutskever, and O.~Vinyals, ``Recurrent neural network regularization,'' \emph{arXiv preprint arXiv:1409.2329}, 2014.

\bibitem{GNN}
F.~Scarselli, M.~Gori, A.~C. Tsoi, M.~Hagenbuchner, and G.~Monfardini, ``The graph neural network model,'' \emph{IEEE transactions on neural networks}, vol.~20, no.~1, pp. 61--80, 2008.

%10
\bibitem{rl_kgs}
Z.~Cui, H.~Chen, L.~Cui, S.~Liu, X.~Liu, G.~Xu, and H.~Yin, ``Reinforced kgs reasoning for explainable sequential recommendation,'' \emph{World Wide Web}, vol.~25, no.~2, pp. 631--654, 2022.

%12
\bibitem{rl_time}
Y.~Zhao, X.~Wang, J.~Chen, Y.~Wang, W.~Tang, X.~He, and H.~Xie, ``Time-aware path reasoning on knowledge graph for recommendation,'' \emph{ACM Transactions on Information Systems}, vol.~41, no.~2, pp. 1--26, 2022.





\bibitem{rs_hgn}
C.~Ma, P.~Kang, and X.~Liu, ``Hierarchical gating networks for sequential recommendation,'' in \emph{Proceedings of the 25th ACM SIGKDD international conference on knowledge discovery \& data mining}, 2019, pp. 825--833.



\bibitem{rl_kg}
H.~Hou and C.~Shi, ``Explainable sequential recommendation using knowledge graphs,'' in \emph{Proceedings of the 5th International Conference on Frontiers of Educational Technologies}, 2019, pp. 53--57.

\bibitem{rl_se}
Y.~Li, H.~Chen, Y.~Li, L.~Li, S.~Y. Philip, and G.~Xu, ``Reinforcement learning based path exploration for sequential explainable recommendation,'' \emph{IEEE Transactions on Knowledge and Data Engineering}, 2023.


\bibitem{rl_mcore}
X.~Li, Y.~Shen, and L.~Chen, ``Mcore: Multi-agent collaborative learning for knowledge-graph-enhanced recommendation,'' in \emph{2021 IEEE International Conference on Data Mining (ICDM)}.\hskip 1em plus 0.5em minus 0.4em\relax IEEE, 2021, pp. 330--339.


\bibitem{ml_gmm}
C.~Stauffer and W.~E.~L. Grimson, ``Adaptive background mixture models for real-time tracking,'' in \emph{Proceedings. 1999 IEEE computer society conference on computer vision and pattern recognition (Cat. No PR00149)}, vol.~2.\hskip 1em plus 0.5em minus 0.4em\relax IEEE, 1999, pp. 246--252.

\bibitem{haarnoja2018soft}
T.~Haarnoja, A.~Zhou, K.~Hartikainen, G.~Tucker, S.~Ha, J.~Tan, V.~Kumar, H.~Zhu, A.~Gupta, P.~Abbeel \emph{et~al.}, ``Soft actor-critic algorithms and applications,'' \emph{arXiv preprint arXiv:1812.05905}, 2018.

\bibitem{afsar2022reinforcement}
M.~M. Afsar, T.~Crump, and B.~Far, ``Reinforcement learning based recommender systems: A survey,'' \emph{ACM Computing Surveys}, vol.~55, no.~7, pp. 1--38, 2022.

\bibitem{ml_transe}
Q.~Ai, V.~Azizi, X.~Chen, and Y.~Zhang, ``Learning heterogeneous knowledge base embeddings for explainable recommendation,'' \emph{Algorithms}, vol.~11, no.~9, p. 137, 2018.

\bibitem{da_amazon}
R.~He and J.~McAuley, ``Ups and downs: Modeling the visual evolution of fashion trends with one-class collaborative filtering,'' in \emph{proceedings of the 25th international conference on world wide web}, 2016, pp. 507--517.

\bibitem{rl_rcenr}
H.~Jiang, C.~Li, J.~Cai, and J.~Wang, ``Rcenr: A reinforced and contrastive heterogeneous network reasoning model for explainable news recommendation,'' in \emph{Proceedings of the 46th International ACM SIGIR Conference on Research and Development in Information Retrieval}, 2023, pp. 1710--1720.


\end{thebibliography}
\end{document}